\documentclass[smallcondensed]{svjour3}

\def\B{{\bf B}}

\def\E{{\mathbb{E}}}

\def\L{{\bf L}}

\def\S{{\bf S}}

\def\T{{\bf T}}

\def\0{{\bf 0}}
\def\1{{\bf 1}}

\def\OM{{\mathcal O}}

\def\Ome{\mbox{\boldmath$\Omega$\unboldmath}}

\def\argmin{\mathop{\rm argmin}}

\def\Lsize{\hbox{\space \raise-2mm\hbox{$\textstyle \L \atop \scriptstyle {m\times 3n}$} \space}}
\def\Ssize{\hbox{\space \raise-2mm\hbox{$\textstyle \S \atop \scriptstyle {m\times 3n}$} \space}}
\def\Osize{\hbox{\space \raise-2mm\hbox{$\textstyle \Ome \atop \scriptstyle {m\times 3n}$} \space}}
\def\Tsize{\hbox{\space \raise-2mm\hbox{$\textstyle \T \atop \scriptstyle {3n\times n}$} \space}}
\def\Bsize{\hbox{\space \raise-2mm\hbox{$\textstyle \B \atop \scriptstyle {m\times n}$} \space}}

\newcommand{\rev}[1]{\textcolor{black}{#1}}

\newcommand{\ip}[2]{\left\langle #1, #2 \right \rangle}

\newcommand{\AName}{FedASMU}
\newcommand{\SName}{FedSSMU}

\usepackage{helvet}  
\usepackage{courier}  
\usepackage[hyphens]{url}  
\usepackage{graphicx} 
\usepackage{caption} 
\DeclareCaptionStyle{ruled}{labelfont=normalfont,labelsep=colon,strut=off} 
\usepackage{amsmath} 
\usepackage{amsfonts} 
\usepackage{amssymb} 
\usepackage{amsthm}
\usepackage{mathrsfs}
\newtheorem{assumption}{Assumption}

\usepackage{xcolor}
\usepackage{subfigure}
\usepackage{booktabs}
\usepackage{multirow}
\usepackage{multicol}
\usepackage{color}
\usepackage{textpos}

\usepackage{algorithm}
\usepackage{algorithmic}

\usepackage{newfloat}
\usepackage{listings}
\floatstyle{ruled}
\newfloat{listing}{tb}{lst}{}
\floatname{listing}{Listing}

\begin{document}

\title{Efficient Federated Learning with Timely Update Dissemination}

\author{Juncheng Jia\footnotemark[4]\footnotemark[5],
        Ji Liu\footnotemark[4]\footnotemark[1]\footnotemark[6],
        Chao Huo\footnotemark[5],
        Yihui Shen\footnotemark[5],
        Yang Zhou\footnotemark[7],
        Huaiyu Dai\footnotemark[8],~and~
        Dejing Dou\footnotemark[9]
}

\date{Received: date / Accepted: date}

\institute{
\footnotemark[1] Corresponding author: jiliuwork@gmail.com.\\
\footnotemark[4]~Equal contribution.\\
\footnotemark[5] J. Jia, C. Huo, Y. Shen are with Soochow University, Suzhou, China. \\
\footnotemark[6] J. Liu is with HiThink Research, Hangzhou China. \\
\footnotemark[7] Y. Zhou is with Auburn University, United States.\\
\footnotemark[8] H. Dai is with North Carolina State University, United States.\\
\footnotemark[9] D. Dou is with Fudan University, Shanghai and BEDI Cloud, Beijing, China. 
}

\authorrunning{Jia and Liu et al.}

\maketitle


\begin{abstract}

Federated Learning (FL) has emerged as a compelling methodology for the management of distributed data, marked by significant advancements in recent years. This paradigm facilitates collaborative model training by utilizing raw data distributed across multiple edge devices. \rev{Compared to traditional machine learning approaches that require centralized data collection, FL offers significant advantages in terms of privacy preservation and the efficient use of edge computing resources.} Nonetheless, the data is typically characterized by non-Independent and Identically Distributed properties, resulting in statistical heterogeneity that can adversely affect model accuracy. Moreover, the disparity in computational and communication capabilities among edge devices (system heterogeneity) further extends the duration of the training process. To address these challenges associated with heterogeneity, we propose an efficient FL approach that capitalizes on additional downlink bandwidth resources to ensure timely update dissemination. Initially, we implement this strategy within an asynchronous framework, introducing the Asynchronous Staleness-aware Model Update (\AName{}), which integrates both server-side and device-side methodologies. On the server side, we present an asynchronous FL system model that employs a dynamic model aggregation technique, which harmonizes local model updates with the global model to enhance both accuracy and efficiency. Concurrently, on the device side, we propose an adaptive model adjustment mechanism that integrates the latest global model with local models during training to further elevate accuracy. Subsequently, we extend this approach to a synchronous context, referred to as \SName{}. Theoretical analyses substantiate the convergence of our proposed methodologies. Extensive experiments, encompassing six models and five public datasets, demonstrate that \AName{} and \SName{} significantly surpass baseline methods in terms of both accuracy (up to 
145.87\%) and efficiency (up to 97.59\%).


\end{abstract}

\section{Introduction}

The rapid expansion of edge devices in recent years has led to the creation of significant volumes of distributed data. Conventional training methods often require this data to be centralized in a server or data center. However, the introduction of regulatory frameworks such as the General Data Protection Regulation (GDPR) \cite{GDPR}, the Consumer Privacy Bill of Rights (CPBR) \cite{Gaff2014}, and the Cybersecurity Law of the People’s Republic of China (CLPR) \cite{CCL} has made these centralized approaches increasingly difficult to implement.
In response, Federated Learning (FL) \cite{mcmahan2021advances,liu2022distributed} has emerged as a viable alternative, allowing for collaborative model training through the transfer of gradients or models rather than raw data. This strategy effectively addresses privacy and security issues linked to direct data transmission while leveraging the computational power of numerous edge devices to develop a global model. FL has demonstrated its versatility across various fields, including computer vision \cite{liu2020fedvision}, natural language processing \cite{liu2021federated}, and healthcare \cite{chen2021fl,silva2020fed,nguyen2022federated}.

Traditional Federated Learning (FL) frameworks typically utilize a parameter server to manage the training process across devices, often relying on a synchronous approach \cite{mcmahan2017communication,liu2023distributed,Jia2023Efficient}. This synchronous training is organized into multiple rounds, each consisting of several essential steps. 
To begin, the server selects a subset of devices for participation \cite{shi2020joint}. It then transmits the current global model to these selected devices. Each device performs local training using its own dataset. Once local training is completed, the devices send their updated models or gradients back to the server. The final step involves the server aggregating the models received from all participating devices to create a new global model, thereby concluding the round once all devices have completed the necessary steps.

Although the synchronous mechanism in FL is effective and relatively easy to implement, its performance can be significantly hindered by the heterogeneity of devices \cite{lai2021oort}. These devices often vary greatly in their computational and communication capabilities \cite{nishio2019client,che2022federated,che2023fast}, as well as in the distribution of their data \cite{mcmahan2017communication,Wang2020Tackling,che2023federated}. For example, some devices may finish local training and model updates quickly, while others may take considerably longer due to constraints such as limited processing power, lower bandwidth, or increased latency—collectively referred to as system heterogeneity.
Moreover, the data stored on each device is frequently non-Independent and Identically Distributed (non-IID), which introduces another layer of complexity known as statistical heterogeneity. This form of heterogeneity can lead to varying local objectives across devices \cite{Wang2020Tackling} and client drift \cite{Karimireddy2020SCAFFOLD,hsu2019measuring}, both of which negatively impact the accuracy of the global model in FL.

Asynchronous FL \cite{xu2021asynchronous,wu2020safa,nguyen2022federated} enables the server to aggregate models that have been uploaded without waiting for slower devices, thereby improving overall efficiency. However, this method may encounter accuracy challenges due to issues such as stale model updates and non-IID data. For instance, when a device uploads a model that is based on an outdated version of the global model, it is possible that the global model has undergone several updates since then. This can lead to the aggregation of the outdated model, effectively reverting the global model to an earlier state and diminishing its accuracy. Furthermore, the asynchronous FL approach may experience convergence difficulties \cite{Su2022How} if the management of staleness is not handled properly \cite{xie2019asynchronous}.

\rev{Current research on FL heterogeneity can be categorized into two main streams: addressing system heterogeneity and tackling statistical heterogeneity. For system heterogeneity, several approaches have been proposed. Device scheduling techniques \cite{shi2020joint,shi2020device,wu2020safa,zhou2022efficient,liu2022multi} aim to select devices based on their computational and communication capacities to balance the load. However, these methods may reduce accuracy due to decreased participation from less capable devices. Other strategies include model pruning \cite{Zhang2022FedDUAP} and dropout \cite{horvath2021fjord}, which can lead to lossy compression and accuracy degradation. Device clustering \cite{Li2022FedHiSyn} and hierarchical architectures \cite{abad2020hierarchical} attempt to group similar devices, but still face efficiency and accuracy challenges in synchronous settings.}
\rev{For statistical heterogeneity, various model aggregation methods have been developed. These include regularization techniques \cite{Li2020FedProx,acar2021federated}, gradient normalization \cite{Wang2020Tackling}, classifier calibration \cite{Luo2021No}, and momentum-based approaches \cite{hsu2019measuring,reddi2021adaptive}. Additionally, contrastive learning \cite{li2021model}, personalization \cite{Sun2021PartialFed,ozkara2021quped}, meta-learning \cite{Khodak2019Adaptive}, and multi-task learning \cite{Smith2017Federated} have been employed to adapt models to non-IID data. However, these methods often lack dynamic adjustment capabilities and are primarily focused on synchronous FL.}
\rev{In the context of asynchronous FL, existing methods typically rely on static polynomial formulas \cite{xie2019asynchronous,Su2022How,chen2021fedsa} or basic attention mechanisms \cite{chen2020asynchronous} for managing staleness. These approaches lack the flexibility to dynamically adjust model importance based on real-time training metrics, such as staleness and loss values. More recent advancements have explored reinforcement learning-based attention \cite{wang2020attention}, meta-learning for staleness management \cite{shen2023federated}, and context-aware aggregation \cite{yi2023fedssa}, but these are still in early stages and not widely adopted.}
\rev{Our work addresses these gaps by introducing a dynamic, staleness-aware model update mechanism that adjusts model importance in real-time, leveraging additional downlink bandwidth for timely update dissemination. This approach not only mitigates the effects of system and statistical heterogeneity but also enhances both accuracy and efficiency in FL systems.}

During our study, we made several key observations. First, we identified the problem of delayed dissemination of updates from local training within both synchronous and asynchronous FL protocols. Additionally, we noted a discrepancy between FL protocols and the properties of bandwidth resources; specifically, FL methods often assume equal data transmission in both directions, despite significant differences in bandwidth capacities. These factors collectively affect overall system efficiency.
To address these challenges, we propose a novel approach that utilizes additional downlink bandwidth resources for timely update dissemination. Specifically, we introduce an innovative Asynchronous Staleness-aware Model Update, denoted as \AName{}. In this framework, we tackle system heterogeneity by designing an asynchronous FL system and implementing a dynamic adjustment method that recalibrates the importance of local models and the global model based on staleness and local loss metrics, aiming to enhance both accuracy and efficiency. \rev{Here, ``importance'' refers to the weighting of the global model in aggregation, dynamically adjusted to balance local and global contributions.} This approach enables devices to adaptively merge fresh global models, thereby reducing the staleness of local models. 
Furthermore, we extend this concept to a synchronous context, referred to as the Synchronous Staleness-aware Model Update (\SName{}) to achieve better accuracy. This paper represents a significant expansion of our previous conference work \cite{liu2024fedasmu}, incorporating the new \SName{} approach, additional theoretical insights, and comprehensive experimental results. The extension includes approximately 35\% new material and additional sections.
The key contributions of this paper are summarized as follows:

\begin{itemize}
    \item \rev{We identify and address the critical issue of delayed model update dissemination in both synchronous and asynchronous FL protocols, which has been underexplored in existing literature. Unlike previous works that primarily focus on either system or statistical heterogeneity, our approach leverages additional downlink bandwidth resources to ensure timely update dissemination, thereby mitigating the inefficiencies caused by delayed updates.}
    
    \item We present an asynchronous staleness-aware model update framework that incorporates methodologies from both the server and client sides. The server-side component modifies the relevance of updated local models and the global model based on staleness and local loss impacts, thus improving accuracy and efficiency. Concurrently, the device-side component updates local models to incorporate fresh global models, thereby reducing staleness and enhancing accuracy.
    
    \item We introduce a synchronous staleness-aware model update framework that adopts similar principles to the asynchronous framework, thereby enhancing performance in synchronous FL settings (new contribution compared with \cite{liu2024fedasmu}).
    
    \item We undertake a comprehensive theoretical analysis to demonstrate the convergence properties of our proposed methods. Additionally, we conduct extensive experiments comparing \AName{} and \SName{} (new contribution compared with \cite{liu2024fedasmu}) with nine state-of-the-art baseline approaches, employing six representative models and five publicly available datasets. The findings illustrate that both \AName{} and \SName{} effectively tackle heterogeneity challenges and significantly surpass the performance of the baseline methods.
\end{itemize}

The structure of the paper is outlined as follows. We begin by discussing the related work in Section \ref{sec:relatedWork}. Next, in Section \ref{sec:systemModel}, we formulate the problem and describe the system model. Section \ref{sec:async} introduces our proposed method for timely update dissemination in the asynchronous context. We then extend this approach to the synchronous scenario in Section \ref{sec:sync}. The results of our experiments are presented in Section \ref{sec:exp}. Finally, we wrap up the paper with conclusions in Section \ref{sec:con}.

\section{Related Work}
\label{sec:relatedWork}

\rev{
Federated learning (FL) is instrumental in advancing knowledge-based systems such as Knowledge-Defined Networking (KDN) \cite{wijesekara2023comprehensive} by facilitating distributed knowledge generation, thereby enhancing scalability and privacy preservation. In KDN, where the knowledge plane processes extensive network data to derive actionable insights, FL enables edge devices to train machine learning models locally, transmitting only model updates rather than raw data. This approach substantially reduces bandwidth requirements, mitigates privacy risks through data localization, and alleviates computational demands on centralized systems. By supporting real-time, privacy-preserving knowledge generation, FL enables applications such as traffic prediction, anomaly detection, and resource allocation, establishing KDN as a cornerstone for self-adaptive and autonomous next-generation intelligent networks \cite{wijesekara2023comprehensive,yuan2024fellas,yuan2024hetefedrec}. Recent scholarly work further underscores FL's significance in knowledge-based systems. Chaudhary et al. \cite{chaudhary2025systematic} provide a comprehensive review of FL systems, emphasizing their efficacy in privacy-preserving distributed machine learning, which is critical for knowledge-based systems like KDN that manage sensitive data. Similarly, Liu et al. \cite{liu2022distributed} offer a detailed survey tracing the evolution from distributed machine learning to FL, highlighting its paradigm shift toward collaborative training without centralized data sharing. This framework directly supports KDN's reliance on distributed data processing for network insights, achieving a robust balance between privacy and system efficiency.
}

\rev{
To address system and statistical heterogeneity challenges in FL, Fog and Edge computing architectures have emerged as key enablers for scalable and resource-efficient FL deployments \cite{xia2021survey,zhao2020federated,wang2019edge,aggarwal2024federated}. These architectures distribute computational resources closer to data sources, reducing latency and enhancing privacy preservation. For instance, Xia et al. \cite{xia2021survey} provide a comprehensive survey of FL in Edge computing, highlighting research problems and solutions related to resource management and privacy. Zhao et al. \cite{zhao2020federated} discuss the fundamental theory and key techniques of FL-enabled intelligent Fog radio access networks, emphasizing Fog computing's role in supporting IoT applications. Wang et al. \cite{wang2019edge} introduce In-Edge AI, which leverages FL to intelligentize mobile Edge computing, caching, and communication, demonstrating enhanced efficiency and scalability. Additionally, Aggarwal et al. \cite{aggarwal2024federated} offer an extensive systematic review of FL in IoT, covering applications and challenges in privacy, resource management, and data management. Furthermore, Hendaoui et al. \cite{hendaoui2025fladen} propose FLADEN, a FL framework for anomaly detection in IoT networks, showcasing practical FL applications in Edge environments. These works collectively underscore the synergy between FL and Fog/Edge computing, providing a foundation for our proposed approach of leveraging additional downlink bandwidth for timely update dissemination in both asynchronous and synchronous FL settings.
}

\rev{
Security and privacy are critical concerns in FL due to its distributed nature, which exposes it to threats such as model poisoning attacks, where malicious clients manipulate updates to degrade the global model, and privacy leakage risks, where sensitive data may be inferred from shared gradients \cite{bhagoji2019analyzing,melis2019exploiting}. Techniques like differential privacy, which adds noise to model updates, and secure aggregation, using cryptographic methods like homomorphic encryption, have been proposed to mitigate these issues \cite{dwork2006differential,bonawitz2016practical}. While these security and privacy challenges are vital, our work primarily focuses on addressing system and statistical heterogeneity through timely update dissemination using additional downlink bandwidth. We plan to explore integrating differential privacy and secure aggregation in future work to enhance the robustness and privacy of our \AName{} and \SName{} frameworks.
}

Numerous FL approaches \cite{mcmahan2017communication,Li2020FedProx,Wang2020Tackling,Karimireddy2020SCAFFOLD,acar2021federated} have been developed to collaboratively train a global model using distributed data from mobile devices. Most of these approaches \cite{bonawitz2019towards} rely on a synchronous mechanism for model aggregation at the server. In this synchronous framework, the server must wait for all selected devices to upload their models before proceeding with aggregation. This process can be inefficient due to the presence of stragglers. The likelihood of encountering stragglers increases with the scale of devices, given the diverse availability and system heterogeneity among devices.

\rev{To address system heterogeneity within the synchronous FL framework, three primary approaches have been widely explored. The first involves client selection and scheduling techniques \cite{shi2020joint,shi2020device,wu2020safa,lai2021oort,nishio2019client,buyukates2021timely}, which strategically choose devices based on their computational and communication capacities to optimize participation and reduce delays. However, these methods often exclude less capable devices, potentially compromising model generalization and accuracy. The second approach employs model compression techniques such as pruning \cite{Zhang2022FedDUAP}, dropout \cite{horvath2021fjord} or parameter-efficient methods\cite{yuan2024hide}, aiming to lighten the computational load on devices. While effective in reducing resource demands, these techniques can introduce lossy compression, leading to degraded model performance. The third approach leverages device clustering \cite{Li2022FedHiSyn} or hierarchical architectures \cite{abad2020hierarchical}, grouping devices with similar capabilities to streamline aggregation. Despite their potential, these methods still struggle with efficiency in highly heterogeneous settings and may not fully address statistical heterogeneity.}


Various model aggregation methods have been proposed to address statistical heterogeneity within the synchronous FL framework. These methods include regularization \cite{Li2020FedProx,acar2021federated}, gradient normalization \cite{Wang2020Tackling}, classifier calibration \cite{Luo2021No}, and momentum-based techniques \cite{hsu2019measuring,reddi2021adaptive}, which aim to adjust local objectives and mitigate accuracy degradation caused by heterogeneous data. Additionally, approaches such as contrastive learning \cite{li2021model}, personalization \cite{Sun2021PartialFed,ozkara2021quped}, meta-learning-based methods \cite{Khodak2019Adaptive}, and multi-task learning \cite{Smith2017Federated} are employed to adapt the global model or local models to non-IID data. Despite their contributions, these methods generally lack the capability to dynamically adjust the importance of diverse models and predominantly focus on the synchronous mechanism.

To address the problem of system heterogeneity, asynchronous FL \cite{xu2021asynchronous,nguyen2022federated} facilitates server-side model aggregation without requiring all devices to be synchronized. Asynchronous FL can proceed either as soon as a model is uploaded from any device \cite{xie2019asynchronous} or when multiple models are accumulated \cite{Nguyen2022FedBuff}. However, the presence of outdated models can revert the global model to a previous state, significantly impairing its accuracy \cite{Su2022How}.
Several approaches have been proposed to enhance the accuracy of asynchronous FL. For example, some methods consider the staleness and divergence of model updates to adjust the importance of uploaded models \rev{\cite{Su2022How,liu2021fedpa}}. However, these methods do not dynamically adapt the weights based on training status, such as loss values. Other methods utilize attention mechanisms and average local training times to adjust model weights \cite{chen2020asynchronous} but do not account for staleness. Additionally, replacing severely stale models with the latest global model \cite{wu2020safa} can mitigate staleness effects but may result in the loss of valuable device-specific information.
Moreover, approaches that use staleness-based polynomial formulas to assign higher weights to fresher models \cite{park2021sageflow,xie2019asynchronous,chen2021fedsa} or leverage model loss values to adjust model importance \cite{park2021sageflow} still rely on static formulas. These static methods are insufficient for dynamically adjusting model importance to minimize loss and improve accuracy.

\rev{Our key observations for the limitations of the existing FL methods are two folded:}
\rev{
\begin{enumerate}
    \item \textbf{Delayed model update dissemination.} In synchronous FL, once fast devices upload their model updates to the server, these updates remain on the server and are not delivered to slower devices until the next round. In asynchronous FL, a device performs local model updates based on an outdated global model, while the global model has already been updated multiple times. In both scenarios, the latest model information is not promptly disseminated to devices, leading to prolonged convergence time.
    \item \textbf{Inefficient downlink bandwidth usage.} The downlink bandwidths of edge devices are typically much smaller than their uplink bandwidths. However, conventional FL methods perform the same amount of data transmission in both directions. As a result, the downlink bandwidth is underutilized during the FL process.
\end{enumerate}
}

Unlike existing approaches, we propose leveraging additional downlink bandwidth resources to disseminate updates in a timely manner.  This strategy helps to mitigate the impact of system heterogeneity. 

\section{System Model and Motivation}
\label{sec:systemModel}

\begin{table} [!t]
\caption{\rev{Summary of main notations}}
\label{tab:summary}
\begin{center}
\begin{tabular}{ |p{0.2\linewidth} | p{0.8\linewidth}| }
\hline
Notation & Definition \\
\hline
$\mathcal{M}$; $m$ & The set of edge devices; the size of $\mathcal{M}$ \\
$\mathcal{D}$; $D$ & The global dataset; the size of $\mathcal{D}$\\
$\mathcal{D}_i$; $D_i$ &  The dataset on Device $i$; the size of $\mathcal{D}_i$\\
$\mathcal{F}(\cdot)$; $\mathcal{F}_i(\cdot)$ & The global loss function; the local loss function on Device $i$ \\
$T$ & The maximum number of global rounds \\
$\mathcal{L}^i$& The maximum number of local epochs on Device $i$ \\
$\tau$& The maximum staleness \\
$\mathcal{T}$ & The constant time period to trigger devices\\
$m'$ & The number of devices to trigger within each time period  \\
$\boldsymbol{w}_t$ & The global model of Version $t$\\
$\boldsymbol{w}_o^i$ & The updated local model from Device $i$ with the original version $o$\\
$\boldsymbol{w}_{o,l}$ & The updated local model with the original version $o$ at local epoch $l$\\
$\boldsymbol{w}_g$ & The fresh global model of Version $g$\\
$\lambda_t^i$, $\sigma_t^i$, $\iota_t^i$ & The control parameters of Device $i$ within the $t$-th local training on the Server\\
$\eta_{\lambda^i}$, $\eta_{\sigma^i}$, $\eta_{\iota^i}$ & The learning rates to update control parameters for Device $i$ on the Server\\
$\alpha^i_t$  & The weight of updated local model from Device $i$ and the $t$-th local training \\
$\beta^i_{t_i}$  & The weight of fresh global model on Device $i$ for the $t_i$-th device-side model aggregation \\
$\gamma^i_{t_i}$, $\upsilon^i_{t_i}$ & The control parameters of Device $i$ for the $t_i$-th device-side model aggregation\\
$\eta_{\gamma^i}$, $\eta_{\upsilon^i}$ & The learning rates to update control parameters for Device $i$ on devices\\
$\eta_i$ & The learning rate on Device $i$ \\
$\eta^i_{RL}$ & The learning rate for the update of RL on Device $i$ \\
$\Theta_t$ & The parameters in the RL model at global round $t$\\
\hline
\end{tabular}
\end{center}
\end{table}

In this section, we define the problem formulation for FL, detail the system model, and explain the motivations driving this research. A summary of the key notations is provided in Table \ref{tab:summary}.

\subsection{System Model}

We examine an FL framework that includes a robust server and \( m \) devices, referred to as \( \mathcal{M} \), working together to collaboratively train a global model.
Each device $i$ stores a local dataset $\mathcal{D}_i = \{\boldsymbol{x}_{i,d} \in \mathbb{R}^s, y_{i,d} \in \mathbb{R} \}_{d=1}^{D_i}$ with $D_i = |\mathcal{D}_i|$ data samples where $\boldsymbol{x}_{i,d}$ is the $d$-th $s$-dimensional input data vector, and $y_{i,d}$ is the label of $\boldsymbol{x}_{i,d}$. The whole dataset is denoted by $\mathcal{D} = \bigcup_{i \in \mathcal{M}} \mathcal{D}_i$ with $D=\sum_{i \in \mathcal{M}} D_i$. Then, the objective of the training process within FL is:
\begin{equation}\label{eq:FLProblem}
\min_{\boldsymbol{w}}\Big\{\mathcal{F}(\boldsymbol{w}) \triangleq \frac{1}{D} \sum_{i \in \mathcal{M}}{D}_i \mathcal{F}_i(\boldsymbol{w}) \Big\}, \tag{$\mathcal{P}1$}
\end{equation}
where $\boldsymbol{w}$ represents the global model, $\mathcal{F}_i(\boldsymbol{w})$ is the local loss function defined as $\mathcal{F}_i(\boldsymbol{w}) \triangleq \frac{1}{D_i}\sum_{\{\boldsymbol{x}_{i,d}, y_{i,d}\} \in \mathcal{D}_i}F(\boldsymbol{w}, \boldsymbol{x}_{i,d}, y_{i,d})$, and $F(\boldsymbol{w}, \boldsymbol{x}_{i,d}, y_{i,d})$ is the loss function to measure the error of the model parameter $\boldsymbol{w}$ on data sample $\{\boldsymbol{x}_{i,d}, y_{i,d}\}$. 

\subsubsection{Synchronous FL}

In synchronous FL, the server initializes a global model and distributes it to $m' (m' \le m) $ devices.
Each selected device uses its local data to train the model for a specified number of epochs. During this phase, the model parameters are updated locally on each client without sharing the data itself.
After local training, devices sends its updated model parameters to the central server.
The server aggregates the received parameters to update the global model. The updated global model is then redistributed to devices, and the process repeats until the model converges or a predefined stopping criterion is met. 

In synchronous FL, all devices must complete their local training and submit their updates before the global model is updated. This ensures that each round of training incorporates contributions from all selected device, leading to more stable and reliable model updates. However, this requirement can also lead to delays, as the system must wait for the slowest client (also known as the ``straggler problem'') to complete its local training before proceeding.

\subsubsection{Asynchronous FL}

Asynchronous FL is a variation of the original FL designed to address some of the limitations of the synchronous approach, particularly the issues related to the ``straggler problem'' and the need for all selected devices to synchronize at each round. In asynchronous FL, devices can independently update and send their local model parameters to the server without waiting for other devices, leading to a more flexible and potentially faster training process. Due to devices updating the global model at varying times, the model parameters submitted to the server may be based on outdated versions of the global model. This ``staleness'' can diminish the effectiveness of each individual update and impede the overall convergence rate of the model.

\subsection{Motivation}



Based on the observations at the end of Section \ref{sec:relatedWork}, in order to address the problem defined in Equation \eqref{eq:FLProblem}, we propose a timely update dissemination approach for both synchronous and asynchronous FL. 
The main idea is to enable the server to promptly disseminate fresh model updates from devices that have completed one round of local training to other devices that are still conducting local training. 
\rev{This method trades increased downlink overhead for higher accuracy and faster training. It assumes downlink bandwidth is typically larger and underutilized than uplink, which is a condition that usually holds. If downlink bandwidth is limited, model compression or transmitting intermediate data (instead of full models) could mitigate the overhead.}

In Section \ref{sec:async}, we first introduce the asynchronous FL with timely update deissemination. We extend the method to the synchronous case in Section \ref{sec:sync}.

\section{Asynchronous Federated Learning with Timely Update Dissemination}
\label{sec:async}

\begin{figure*}[!t]
\centering
\includegraphics[width=1\linewidth]{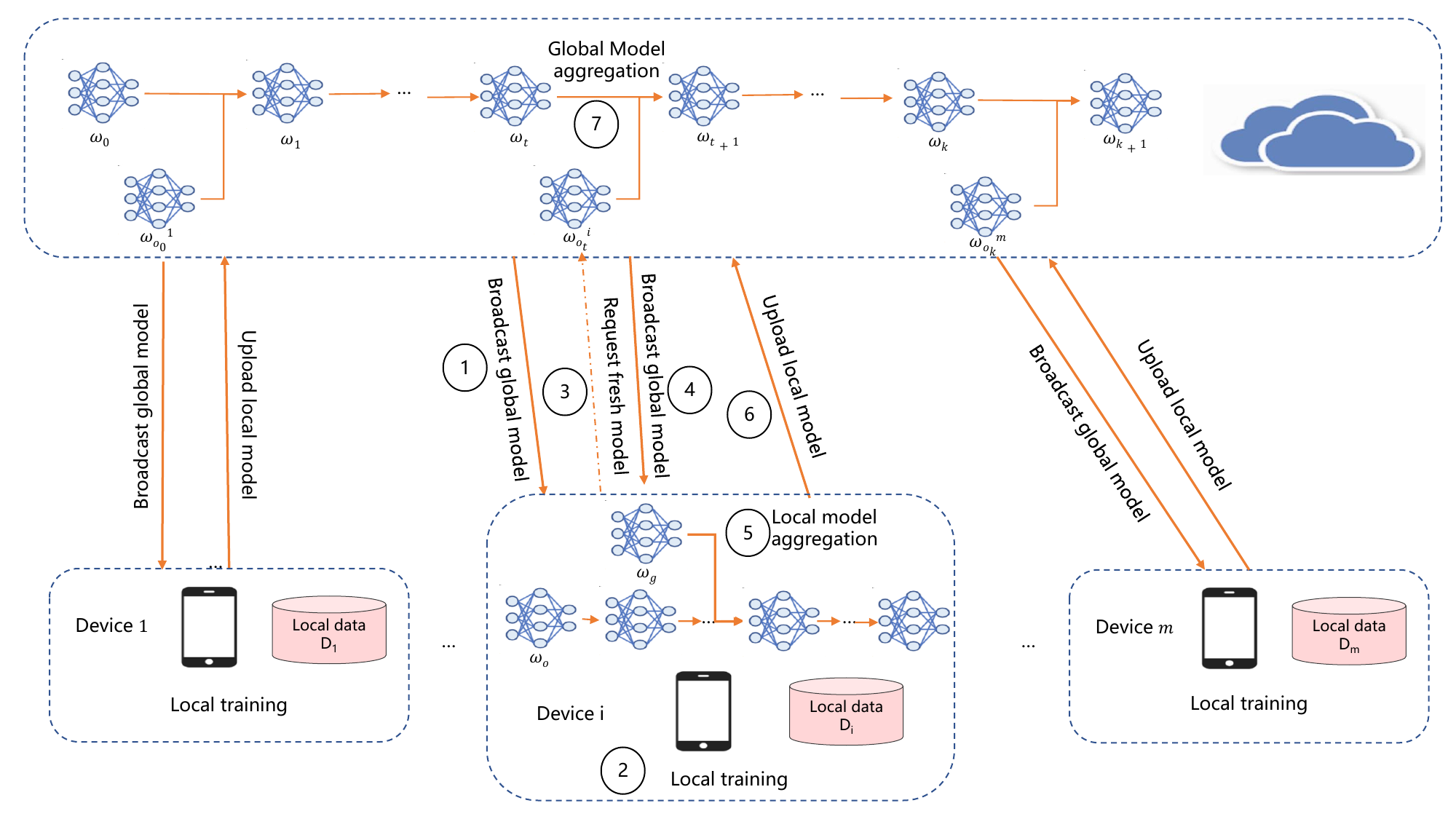}
\caption{The system model of \AName{}.}
\label{fig:system}
\end{figure*}

In this section, we first introduce the overall method of asynchronous federated learning with timely  update dissemination (\AName{}). Then we describe the staleness-aware model update. Finally, we provide convergence analysis.

\subsection{Overall Method}

The overall process of \AName{} is depicted in Figure \ref{fig:system}. The server initiates local training on \( m' \) devices at regular intervals, represented by \( \mathcal{T} \). This training takes place over multiple global rounds, starting with the global model at version 0, which increments by 1 after each round.
Each global round involves seven distinct steps:
\begin{itemize}
    \item Step \textcircled{1}: The server randomly selects and activates \( m' \) eligible devices, where \( m' \leq m \), and transmits the current global model \( \boldsymbol{w}_o \) to these devices.
    \item Step \textcircled{2}: Each chosen device performs local training using its respective dataset.
    \item Step \textcircled{3}: While engaging in local training, Device \( i \) may request an updated global model to mitigate any potential staleness, as the global model could be modified during this period.
    \item Step \textcircled{4}: If the global model \( \boldsymbol{w}_g \) has been revised (i.e., \( g > o \)), the server will provide the updated version to the requesting device.
    \item Step \textcircled{5}: Upon receiving the updated global model, the device integrates it with its current local model to form a new model, which it will continue training.
    \item Step \textcircled{6}: After completing local training, the device sends its local model back to the server.
    \item Step \textcircled{7}: The server combines the latest global model \( \boldsymbol{w}_t \) with the local model \( \boldsymbol{w}^i_o \) from Device \( i \). The staleness of the local model is computed as \( \tau_i = t - o + 1 \), where \( t \) is the current global round. If the staleness \( \tau_i \) surpasses a defined threshold \( \tau \), the local model is discarded to prevent degrading the global model's accuracy with outdated information. This approach ensures that staleness remains within acceptable bounds, supporting convergence \cite{ho2013more}.
\end{itemize}

\rev{
To address the inefficient usage of downlink bandwidth identified in Section~\ref{sec:systemModel}, FedASMU leverages additional downlink bandwidth resources to facilitate timely dissemination of global model updates. Specifically, during local training, devices can request the latest global model from the server (Step \textcircled{3}), which is transmitted using the available downlink capacity (Step \textcircled{4}). This approach ensures that devices receive fresher models, reducing staleness and enhancing training efficiency. Experimental results in Section~\ref{sec:exp} demonstrate that this strategy reduces training time by up to 97.59\% compared to baseline methods, as detailed in Section~\ref{sec:expResults}.
}

In the subsequent part of this section, we present our dynamic staleness-aware model aggregation technique used on the server (Step \textcircled{7}), as well as the adaptive local model adjustment strategy applied on the devices (Steps \textcircled{3} and \textcircled{5}).

\subsection{Dynamic Model Update on the Server}
\label{subsec:MAServer}

In this subsection, we introduce a dynamic staleness-aware model update strategy for the server. Upon receiving the uploaded model \( \boldsymbol{w}_o^i \) from Device \( i \), which corresponds to version \( o \), the server updates the current global model \( \boldsymbol{w}_t \) according to the following equation:
\begin{equation}
\label{eq:serverAggregation}
\boldsymbol{w}_{t+1} = (1-\alpha^i_t) \boldsymbol{w}_{t} + \alpha^i_t \boldsymbol{w}_o^i,
\end{equation}
where $\alpha^i_t$ represents the weight or importance assigned to the uploaded model from Device $i$ during global round $t$, which plays a crucial role in influencing the accuracy of the aggregated global model \cite{xie2019asynchronous}. 

Next, we reformulate the problem defined in Equation \eqref{eq:FLProblem} as a bi-level optimization problem \cite{Bard1998PracticalBO}:
\begin{align}
\min_{\boldsymbol{w},\boldsymbol{A}}\Big\{&\mathcal{F}(\boldsymbol{w}, \boldsymbol{A}) \triangleq \frac{1}{|D|} \sum_{i \in \mathcal{M}}|\mathcal{D}_i| \mathcal{F}_i(\boldsymbol{w}(\boldsymbol{A})) \Big\}, \nonumber\\
&\boldsymbol{w}(\boldsymbol{A}) = (1-\alpha^i_t) \boldsymbol{w}_{t} + \alpha^i_t \boldsymbol{w}_o^i, \alpha^i_t \in \boldsymbol{A}, \nonumber
\\
\textit{s.t.} &~\boldsymbol{w}^i_o = \argmin_{\boldsymbol{w}^i_o}\mathcal{F}_i(\boldsymbol{w}^i_o)~\forall~i \in \mathcal{M}, \label{eq:Problem2} \tag{$\mathcal{P}2$}\\ 
&~\boldsymbol{A} = \argmin_{\boldsymbol{A}}\mathcal{F}(\boldsymbol{w}, \boldsymbol{A}), \label{eq:Problem3} \tag{$\mathcal{P}3$}
\end{align}
where $\boldsymbol{A} = {\alpha_t^1, \alpha_t^2, \dots, \alpha_t^m}$ represents the set of importance values assigned to the models uploaded by devices. Problem \ref{eq:Problem2} focuses on minimizing the local loss function, which is further discussed in Section \ref{subsec:MADevice}.

Inspired by \cite{xie2019asynchronous}, we propose using a dynamic polynomial function to represent $\alpha^i_t$, as defined in Equation \eqref{eq:alpha}, for the optimization problem outlined in Problem \ref{eq:Problem3}:
\begin{equation}
\begin{split}
\label{eq:alpha}
\xi^i_t(o) &= \frac{\lambda^i_t}{\sqrt{t}(t - o +1)^{\sigma^i_t}} + \iota^i_t, \\
\alpha^i_t(o) &= \frac{\mu_{\alpha} \xi^i_t(o)}{1 + \mu_{\alpha} \xi^i_t(o)},
\end{split}
\end{equation}
where $\mu_{\alpha}$ is a hyper-parameter, $t - o + 1$ denotes the staleness, $t$ represents the current version of the global model, and $o$ is the version that Device $i$ received prior to local training. The terms $\lambda^i_t$, $\sigma^i_t$, and $\iota^i_t$ are control parameters for Device $i$ during the $t$-th global round. These parameters are dynamically adjusted according to Equation \eqref{eq:control} to minimize the global model's loss.
\begin{align}
\label{eq:control}
\lambda^i_t &= \lambda^i_{o-1} - \eta_{\lambda^i}\nabla_{\lambda^i_{o-1}} \mathcal{F}(\boldsymbol{w}_{o}), \nonumber\\
\sigma^i_t &= \sigma^i_{o-1} - \eta_{\sigma^i}\nabla_{\sigma^i_{o-1}} \mathcal{F}(\boldsymbol{w}_{o}), \\
\iota^i_t &= \iota^i_{o-1} - \eta_{\iota^i}\nabla_{\iota^i_{o-1}} \mathcal{F}(\boldsymbol{w}_{o}), \nonumber
\end{align}
where $\eta_{\lambda^i}$, $\eta_{\sigma^i}$, and $\eta_{\iota^i}$ represent the corresponding learning rates for dynamic adjustment, $\nabla_{\lambda^i_{o-1}}\mathcal{F}(\boldsymbol{w}_{o})$, $\nabla_{\sigma^i_{o-1}}\mathcal{F}(\boldsymbol{w}_{o})$, and $\nabla_{\iota^i_{o-1}}\mathcal{F}(\boldsymbol{w}_{o})$ correspond to the respective partial derivatives of the loss function. 

Let $\boldsymbol{w}_{o'}^i$ denote the local model corresponding to the $(o - 1)$-th server-side model aggregation. After aggregating the local model $\boldsymbol{w}_{o'}^i$, the global model is updated to version $o$.
\begin{equation*}
\begin{aligned}
\nabla_{\lambda_{o-1}^i} \mathcal{F}(\boldsymbol{w}_{o}) &= \left(\frac{\partial \mathcal{F}(\boldsymbol{w}_{o})}{\partial \boldsymbol{w}_{o}}\right)^\mathrm{T} \frac{\partial \boldsymbol{w}_{o}}{\partial \lambda_{o-1}^i} \\
&\approx \left(\frac{\partial \mathcal{F}_i(\boldsymbol{w}_{o})}{\partial \boldsymbol{w}_{o}}\right)^\mathrm{T} \frac{\partial \boldsymbol{w}_{o}}{\partial \lambda_{o-1}^i}\\
&= g_i^\mathrm{T}(\boldsymbol{w}_{o}) \frac{\partial \boldsymbol{w}_{o}}{\partial \lambda_{o-1}^i},  
\end{aligned}
\end{equation*}
where the symbol $\approx$ denotes the approximation of the global partial deviation of $\boldsymbol{w}_{o}$ by that on Device $i$. 
\begin{equation*}
\begin{aligned}
g_i^\mathrm{T}(\boldsymbol{w}_{o}) \approx \frac{\boldsymbol{w}_{o}^i - \boldsymbol{w}_{o}}{\eta_i \mathcal{L}},
\end{aligned}
\end{equation*}
where $\boldsymbol{w}_{o}^i$ is the updated local model, and $\boldsymbol{w}_{o}$ is the original global model to generate $\boldsymbol{w}_{o}^i$. The calculation of $g_i^\mathrm{T}(\boldsymbol{w}_{o})$ does not incur extra communication. 
\begin{equation*}
\begin{aligned}
\boldsymbol{w}_{o} &= (1-\alpha^i_{o-1})\boldsymbol{w}_{o-1}+\alpha^i_{o-1}\boldsymbol{w}_{o'}^i\\
&=\boldsymbol{w}_{o-1} + \alpha^i_{o-1}(\boldsymbol{w}_{o'}^i - \boldsymbol{w}_{o-1}),
\end{aligned}
\end{equation*}
where $\boldsymbol{w}_{o-1}^i$ and $\boldsymbol{w}_{o-1}$ are independent with $\lambda^i_{o-1}$. Thus, we have:
\begin{equation*}
\begin{aligned}
\frac{\partial \boldsymbol{w}_{o}}{\partial \lambda_{o-1}^i} &= \frac{\partial(\boldsymbol{w}_{o-1} + \alpha^i_{o-1}(\boldsymbol{w}_{o'}^i - \boldsymbol{w}_{o-1}))}{\partial \lambda_{o-1}^i}\\
&= (\boldsymbol{w}_{o'}^i - \boldsymbol{w}_{o-1})\frac{\partial \alpha^i_{o-1}}{\partial \lambda_{o-1}^i}.
\end{aligned}
\end{equation*}
After elaborating $\alpha^i_{o-1}$, we have:
\begin{equation*}
\begin{aligned}
\frac{\partial \alpha^i_{o-1}}{\partial \lambda_{o-1}^i} &= \frac{\partial(1 - (1 + \mu_{\alpha} \xi^i_{o-1})^{(-1)})}{\partial \lambda_{o-1}^i}\\
&= \frac{\mu_{\alpha}}{(1+\mu_{\alpha} \xi^i_{o-1})^{2}}\frac{\partial \xi^i_{o-1}}{\partial \lambda_{o-1}^i}\\
&= \frac{\mu_{\alpha}}{(1+\mu_{\alpha} \xi^i_{o-1})^{2}}\frac{1}{\sqrt{o-1}(o - o')^{\sigma^i_{o-1}}},
\end{aligned}
\end{equation*}
where $\xi^i_{o-1}$ represents $\xi^i_{o-1}(o')$, with $o'$ denoting the version of the original global model used to generate the updated local model $\boldsymbol{w}_{o'}^i$ during the $(o-1)$-th global round.
Finally, the partial deviation of the loss function with respect to $\lambda_{o-1}^i$ can be calculated as Equation\eqref{eq:dlambda}:
\begin{equation}
\begin{aligned}
\label{eq:dlambda}
\nabla_{\lambda^i_{o-1}} \mathcal{F}(\boldsymbol{w}_{o}) \approx \frac{\mu_{\alpha}(\boldsymbol{w}_{o}^i - \boldsymbol{w}_{o})(\boldsymbol{w}_{o'}^i - \boldsymbol{w}_{o-1})}{\eta_i \mathcal{L} \sqrt{o-1}(1+\mu_{\alpha} \xi^i_{o-1})^{2}(o - o')^{\sigma^i_{o-1}}}.
\end{aligned}
\end{equation}
Similarly, the partial deviation of the loss function with respect to $\sigma_{o-1}^i$ and $\iota_{o-1}^i$ can be derived as Equation\eqref{eq:dsigmaiota}:
\begin{equation}
\begin{aligned}
\label{eq:dsigmaiota}
\nabla_{\sigma^i_{o-1}} \mathcal{F}(\boldsymbol{w}_{o}) &\approx \frac{\mu_{\alpha}\ln(\sigma^i_{o-1})(\boldsymbol{w}_{o} - \boldsymbol{w}_{o}^i)(\boldsymbol{w}_{o'}^i - \boldsymbol{w}_{o-1})}{\eta_i \mathcal{L} \sqrt{o-1}(1+\mu_{\alpha} \xi^i_{o-1})^{2}(o - o')^{\sigma^i_{o-1}}},\\
\nabla_{\iota^i_{o-1}} \mathcal{F}(\boldsymbol{w}_{o}) &\approx \frac{\mu_{\alpha}(\boldsymbol{w}_{o}^i - \boldsymbol{w}_{o})(\boldsymbol{w}_{o'}^i - \boldsymbol{w}_{o-1})}{\eta_i \mathcal{L}(1+\mu_{\alpha} \xi^i_{o-1})^{2}}.
\end{aligned}
\end{equation}

\begin{figure}[t]
\begin{algorithm}[H]
\caption{\AName{} on the Server}
\label{alg:serverAggregation}
\begin{algorithmic}[1]
\REQUIRE  \quad \\
$T$: The maximum number of global rounds \\
$m'$: The number of devices to be triggered \\
$\tau$: The predefined staleness limit \\
$\mathcal{T}$: The constant time period to trigger devices \\
$\boldsymbol{w}_0$: The initial global model \\
$\lambda_{0}$, $\sigma_{0}$, $\iota_{0}$: The initial control parameters\\
$\eta_{\lambda^i}$, $\eta_{\sigma^i}$, $\eta_{\iota^i}$: The learning rates for the dynamic adjustment
\ENSURE   \quad \\
$\boldsymbol{w}_{T}$: The global model at Round $T$ \\
\rev{
\STATE \textbf{Description}: This algorithm outlines the server-side procedure for FedASMU. It periodically triggers devices, receives local models, and aggregates them with the global model while managing staleness.
}
\WHILE{The training is not finished (in parallel)} \label{alg:threadBegin}
    \IF{Should trigger new devices}
        \STATE Trigger and broadcast the global model to $m'$ devices for parallel local training \quad \rev{// Initiates local training on selected devices}
        \STATE Sleep $\mathcal{T}$ \quad \rev{// Pauses until the next trigger period}
    \ENDIF
\ENDWHILE \label{alg:threadEnd}
\FOR{$t$ in $\{1, 2, ..., T\}$}
    \STATE Receive $\boldsymbol{w}^i_o$ \quad \rev{// Collects local model updates from devices} \label{alg:receive}
    \IF{$t - o + 1 > \tau$} \label{alg:verify}
        \STATE Discard $\boldsymbol{w}^i_o$ and continue \quad \rev{// Discards outdated models exceeding staleness limit} \label{alg:discard}
    \ELSE
        \STATE Update $\lambda_t^i$, $\sigma_t^i$, $\iota_t^i$ according to Equation \eqref{eq:control} \quad \rev{// Dynamically adjusts control parameters for aggregation} \label{alg:updateControl}
        \STATE Update $\alpha^i_t$ utilizing Equation \eqref{eq:alpha} \quad \rev{// Calculates aggregation weights for local models} \label{alg:updateAlpha}
        \STATE Update $\boldsymbol{w}_t$ exploiting Equation \eqref{eq:serverAggregation} \quad \rev{// Aggregates local models into the global model} \label{alg:updateModel}
    \ENDIF
\ENDFOR
\rev{
\STATE \textbf{Return} $\boldsymbol{w}_{T}$ \quad // Returns the final global model
}
\end{algorithmic}
\end{algorithm}
\end{figure}

The server’s model aggregation algorithm for \AName{} is detailed in Algorithm \ref{alg:serverAggregation}. A separate thread is responsible for periodically activating \( m' \) devices when the number of devices engaged in training falls below a specified threshold (Lines \ref{alg:threadBegin} - \ref{alg:threadEnd}). Upon the receipt of the model \(\boldsymbol{w}^i_o\) (Line \ref{alg:receive}), the server verifies whether this uploaded model falls within the acceptable staleness limits (Line \ref{alg:verify}). 
If the model does not meet the staleness criteria, it is discarded (Line \ref{alg:discard}). Conversely, if it is valid, the server proceeds to update the control parameters \(\lambda_t^i\), \(\sigma_t^i\), and \(\iota_t^i\) according to the specifications in Equation \eqref{eq:control} (Line \ref{alg:updateControl}), and it computes \(\alpha^i_t\) as outlined in Equation \eqref{eq:alpha} (Line \ref{alg:updateAlpha}). Subsequently, the global model is updated by the server (Line \ref{alg:updateModel}).

\subsection{Adaptive Model Update on Devices}
\label{subsec:MADevice}

In this subsection, we detail the local training procedure that incorporates an adaptive local model adjustment technique on devices to tackle Problem \ref{eq:Problem2}.

Once selected for local training, Device $i$ uses the global model $\boldsymbol{w}_{o}$ from the server as the initial local model $\boldsymbol{w}_{o,0}$.
Device $i$ updates the local model based on the local dataset $\mathcal{D}_i$ with the Stochastic Gradient Descent (SGD) approach \cite{robbins1951stochastic}, as defined in Equation \eqref{eq:localSGD}. 
\begin{equation}
\label{eq:localSGD}
   \boldsymbol{w}_{o,l} = \boldsymbol{w}_{o,l-1} - \eta_i \nabla \mathcal{F}_{i}(\boldsymbol{w}_{o,l-1}, \zeta_{l-1}), \ \  \zeta_{l-1} \sim \mathcal{D}_i ,
\end{equation}
where $o$ denotes the global model's version, $l$ is the count of local epochs, $\eta_i$ is Device $i$'s learning rate, and $\nabla \mathcal{F}_{i}(\cdot)$ is the gradient derived from a mini-batch $\zeta_{l-1}$ sampled from $\mathcal{D}_i$.

To minimize the discrepancy between the global model and the local model, we propose merging the updated global model with the local model during the local training process. Since the global model may be updated frequently during local training, incorporating the latest global model can effectively reduce this gap. However, determining the optimal time to request and the appropriate weights for merging the fresh global model is complex. In this section, we first introduce a Reinforcement Learning (RL) method to select an appropriate time slot for devices to request. We then detail the adaptive device-side model update method.

\subsubsection{Request Time Slot Selection}
\label{subsec:slot_selection}

We propose an RL-based intelligent request time slot selector to determine the optimal time for devices to request a new version of global model from the server. 
To reduce communication overhead, we assume that only one updated global model is received during local training. On one hand, making a request too early may result in the server having conducted only a few updates, which can lead to significant staleness in the final local model. On the other hand, if the request is submitted too late, the local update might fail to take advantage of the updated global model, which can result in reduced accuracy. Therefore, it is essential to choose a suitable time frame for making the request.

The intelligent time slot selector comprises a server-side meta model and device-side models. Initially, the meta model produces a time slot decision for every device, which is then updated upon the completion of the device's first local training session. This initial time slot serves as the basis for the local model, which undergoes updates during subsequent training sessions to establish a personalized optimal time slot for requesting the latest global model. For the meta model, we employ a Long Short-Term Memory (LSTM) network with a fully connected layer, while each local model utilizes a $\mathcal{Q}$-learning approach \cite{WatkinsD1992Technical}. Both models calculate the probability associated with each time slot, and we implement the $\epsilon$-greedy strategy \cite{xia2015online} for selection purposes.

\begin{figure}[t]
\begin{algorithm}[H]
\caption{\AName{} on Devices}
\label{alg:device_passive}
\begin{algorithmic}[1]
\REQUIRE  \quad \\
$t$: The number of the meta model update \\
$t_i$: The number of device-side model aggregation \\
$\mathcal{L}^i$: The maximum number of epochs on Device $i$ \\
$\boldsymbol{w}_o$: The original global model with Version $o$ \\
$\boldsymbol{w}_g$: The fresh global model with Version $g$\\
$\theta^i_{t-1}$: The parameters of the meta model \\
$\gamma^i_{t_i-1}$, $\upsilon^i_{t_i-1}$: The control parameters\\
\ENSURE \quad \\
$\boldsymbol{w}^i_{o,\mathcal{L}}$: The trained local model \\
\rev{
\STATE \textbf{Description}: This algorithm details the device-side training process in FedASMU. It integrates local training with periodic updates from the global model to reduce staleness.
}
\STATE $l^* \leftarrow$ Generate a time slot using $\theta^i_{t-1}$ or $\mathcal{H}_{t_i-1}^i$ \quad \rev{// Determines when to request a fresh global model} \label{alg:select}
\STATE $\boldsymbol{w}^i_{o,0} \leftarrow \boldsymbol{w}_o$ \quad \rev{// Initializes local model with global model}
\FOR{$l$ in $\{1, 2, ..., \mathcal{L}^i\}$}
    \IF{$l = l^*$}
        \STATE Send a request for a fresh global model to server \quad \rev{// Requests updated global model to reduce staleness} \label{alg:send}
        \STATE Receive $\boldsymbol{w}_g$ \quad \rev{// Obtains the fresh global model} \label{alg:wait}
    \ENDIF
    \IF{$\boldsymbol{w}_g \neq \boldsymbol{w}_o$} \label{alg:receiveFresh}
        \STATE $\beta^i_{t_i-1} \leftarrow$ Calculation based on Equation \eqref{eq:betaUpdate} \quad \rev{// Computes aggregation weight for model merging} \label{alg:betaUpdate}
        \STATE Update $\boldsymbol{w}_{o,l-1}$ with Equation \eqref{eq:deviceAggregation} \quad \rev{// Aggregates local and fresh global models} \label{alg:deviceAggregation}
        \STATE Update $\gamma^i_{t_i}$ and $\upsilon^i_{t_i}$ with Equation \eqref{eq:controlDevice} \quad \rev{// Updates device-side control parameters} \label{alg:controlDeviceUpdate}
        \STATE $\mathcal{R} \leftarrow loss_{o,l}^{b,i} - loss_{o,l}^{a,i}$ \quad \rev{// Calculates reward based on loss difference} \label{alg:rewardUpdate}
        \STATE $b_{t_i} \leftarrow (1-\rho)b_{t_i-1} + \rho \mathcal{R}$ \quad \rev{// Updates bias term for reinforcement learning} \label{alg:btUpdate}
        \STATE Update $\theta_{t}$ or $\mathcal{H}_{t_i}^i$ with $\mathcal{R}$ \quad \rev{// Updates meta model or history with reward} \label{alg:rlupdate}
    \ENDIF
    \STATE Update $\boldsymbol{w}_{o,l}$ based on Equation \eqref{eq:localSGD} \quad \rev{// Performs local stochastic gradient descent update} \label{alg:SGDUpdate}
\ENDFOR
\rev{
\STATE \textbf{Return} $\boldsymbol{w}^i_{o,\mathcal{L}}$ \quad // Returns the trained local model
}
\end{algorithmic}
\end{algorithm}
\end{figure}

For local training, the reward $\mathcal{R}$ is defined as the change in loss value before and after model merging on a device. Specifically, we denote $loss_{o,l^*}^{b,i}$ as the loss value of $\mathcal{F}_{i}(\boldsymbol{w}_{o,l^*}, \zeta_{l^*})$ before merging the new global model according to the request after $l^*$ local epochs, and denote $loss_{o,l^*}^{a,i}$ as that after merging. Then we have $\mathcal{R} = loss_{o,l^*}^{b,i} - loss_{o,l^*}^{a,i}$. The LSTM model is updated using Equation \eqref{eq:rlupdate} after the initial aggregation \cite{Zoph2017Neural},
\begin{equation}\label{eq:rlupdate}\small
    \theta_t = \theta_{t - 1} + \eta_{RL} \sum_{l = 1}^{ \mathcal{L}} \nabla_{\theta_{t - 1}} \log P(\mathcal{s}_l|\mathcal{s}_{(l-1):1}; \theta_{t - 1})(\mathcal{R} - b_t),
\end{equation}
where $\theta_t$ is the meta model after its $t$-th update, $\eta_{RL}$ is the learning rate of RL training, $\mathcal{L}$ is the maximum number of local epochs, $\mathcal{s}_l$ denotes the requesting decision after the $l$-th local epoch, and $b_t$ is used to control the bias of the model. The model undergoes pre-training using historical data and is subsequently updated in real time during the training process of \AName{} on each device. The $\mathcal{Q}$-learning method establishes a mapping \(\mathcal{H}^i\) between decisions and rewards for Device \(i\). This mapping is revised based on a weighted average that incorporates both historical values and the current reward \cite{dietterich2000hierarchical}, as demonstrated in Equation \eqref{eq:qupdate},
\begin{align}
\label{eq:qupdate}
    &\mathcal{H}^i_{t_i}  (l^*_{t_i - 1}, a_{t_i - 1})  = \mathcal{H}^i_{t_i-1}(l^*_{t_i - 1}, a_{t_i - 1})  \nonumber \\
    & + \phi (\mathcal{R} + \psi \max_a{\mathcal{H}^i_{t_i-1}(l^*_{t_1}, a) - \mathcal{H}^i_{t_i-1}(l^*_{t_i - 1}, a_{t_i - 1})}),
\end{align}
where $a_{t_i - 1}$ denotes the action, $l^*_{t_i - 1}$ is the number of local epochs to send the request within the $(t_i - 1)$-th device-side model aggregation, $\phi$ and $\psi$ are hyper-parameters. The action space for $a_{t_i - 1}$ is $\{add, stay, minus\}$, where $add$ means adding 1 epoch to $l^*_{t_i - 1}$ ($l^*_{t_i} = l^*_{t_i - 1} + 1$), $stay$ means staying with the same epoch ($l^*_{t_i} = l^*_{t_i - 1} + 1$), and $minus$ means removing 1 epoch from $l^*_{t_i - 1}$ ($l^*_{t_i} = l^*_{t_i - 1} - 1$).

\subsubsection{Device-Side Model Aggregation}
\label{subsec:device_agg}

When receiving a fresh global model $\boldsymbol{w}_g$, Device $i$ merges the fresh global model with its current local model $\boldsymbol{w}_{o,l}^b$ with Equation \eqref{eq:deviceAggregation}.
\begin{equation}
\label{eq:deviceAggregation}
\boldsymbol{w}_{o,l}^a = (1-\beta^i_{t_i-1}) \boldsymbol{w}_{o,l}^b + \beta^i_{t_i-1} \boldsymbol{w}_g,
\end{equation}
where $\beta^i_{t_i - 1}$ is the weight of the fresh global model on Device $i$ at the $(t_i - 1)$-th device-side model aggregation. Equation \eqref{eq:deviceAggregation} differs from Equation \eqref{eq:serverAggregation} as the received fresh global model corresponds to a higher global version. $\beta^i_{t - 1}$ is calculated with Equation \eqref{eq:betaUpdate},
\begin{equation}\label{eq:betaUpdate}
\begin{split}
\phi^i_{t_i-1}(g, o) &= \frac{\gamma^i_{t_i-1}}{\sqrt{g}}(1-\frac{\upsilon^i_{t_i-1}}{\sqrt{g - o + 1}}),\\
\beta^i_{t_i-1}(g, o) &= \frac{\mu_{\beta} \phi^i_{t_i-1}(g, o)}{1 + \mu_{\beta} \phi^i_{t_i-1}(g, o)},
\end{split}    
\end{equation}
where $\mu_{\beta}$ is a hyper-parameter, $\gamma^i_{t_i-1}$ and $\upsilon^i_{t_i-1}$ are control parameters adjusted with Equation \eqref{eq:controlDevice},
\begin{align}
\label{eq:controlDevice}
\gamma^i_{t_i} &= \gamma^i_{t_i-1} - \eta_{\gamma^i}\nabla_{\gamma^i_{t_i-1}} \mathcal{F}_i(\boldsymbol{w}_{o,l}^b, \zeta_{l-1}), \nonumber\\
\upsilon^i_{t_i} &= \upsilon^i_{t_i-1} - \eta_{\upsilon^i}\nabla_{\upsilon^i_{t_i-1}} \mathcal{F}_i(\boldsymbol{w}_{o,l}^b, \zeta_{l-1}), \zeta_{l-1} \sim \mathcal{D}_i.
\end{align}
For Equation \eqref{eq:controlDevice}, $\eta_{\gamma^i}$ and $\eta_{\upsilon^i}$ are learning rates for $\gamma^i_{t_i}$ and $\upsilon^i_{t_i}$. Besides, we have the partial deviation of the loss function in terms of $\gamma^i_{t_i-1}$: 
\begin{equation*}
\begin{aligned}
\nabla_{\gamma^i_{t_i-1}} \mathcal{F}_i(\boldsymbol{w}_{o,l}^b, \zeta_{l-1}) &= \left(\frac{\partial \mathcal{F}_i(\boldsymbol{w}_{o,l}^b, \zeta_{l-1})}{\partial \boldsymbol{w}_{o,l}^b} \right)^\mathrm{T} \frac{\partial \boldsymbol{w}_{o,l}^b}{\partial \gamma_{t_i-1}^i} \\
&= g_{o,l}^\mathrm{T}(\boldsymbol{w}_{o,l}^b) \frac{\partial \boldsymbol{w}_{o,l}^b}{\partial \gamma_{t_i-1}^i},  \\
\end{aligned}
\end{equation*}
where $g_{o,l}^\mathrm{T}(\boldsymbol{w}_{o,l}^b)$ is the local gradient on Device $i$ with $\boldsymbol{w}_{o,l}^b$ and $\zeta_{l-1}$. Meanwhile, we have
\begin{equation*}
\begin{aligned}
\boldsymbol{w}_{o,l}^b &= (1-\beta^i_{t_i-1}) \boldsymbol{w}_{o,l}^a + \beta^i_{t_i-1} \boldsymbol{w}_g \\
&= \boldsymbol{w}_{o,l}^a + \beta^i_{t_i-1}(\boldsymbol{w}_g - \boldsymbol{w}_{o,l}^a),
\end{aligned}
\end{equation*}
where $\boldsymbol{w}_g$ and $\boldsymbol{w}_{o,l}^a$ are independent with $\gamma_{t_i-1}^i$. Then, we have:
\begin{equation*}
\begin{aligned}
\frac{\partial \boldsymbol{w}_{o,l}^b}{\partial \gamma_{t_i-1}^i} &= \frac{\partial(\boldsymbol{w}_{o,l}^a + \beta^i_{t_i-1}(\boldsymbol{w}_g - \boldsymbol{w}_{o,l}^a))}{\partial \gamma_{t_i-1}^i} \\
&= (\boldsymbol{w}_g - \boldsymbol{w}_{o,l}^a)\frac{\partial \beta^i_{t_i-1}}{\partial \gamma_{t_i-1}^i}.
\end{aligned}
\end{equation*}
After elaborating $\beta^i_{t_i-1}$, we have:
\begin{equation*}
\begin{aligned}
\frac{\partial \beta^i_{t_i-1}}{\partial \gamma_{t_i-1}^i} &= \frac{\partial(1-(1 + \mu_{\beta} \phi^i_{t_i-1})^{-1})}{\partial \gamma_{t_i-1}^i} \\
&= \frac{\mu_{\beta}}{(1 + \mu_{\beta} \phi^i_{t_i-1})^{2}}\frac{\partial \phi^i_{t_i-1}}{\partial \gamma_{t_i-1}^i} \\
&= \frac{\mu_{\beta}}{\sqrt{g}(1 + \mu_{\beta} \phi^i_{t_i-1})^{2}}(1-\frac{\upsilon^i_{t_i-1}}{\sqrt{g - o + 1}}),
\end{aligned}
\end{equation*}
where $\phi^i_{t_i-1}$ represents $\phi^i_{t_i-1}(g,o)$. Finally, we can calculate the partial deviation of the loss function in terms of $\gamma_{t_i-1}^i$:
\begin{equation*}
\begin{aligned}
&\nabla_{\gamma^i_{t_i-1}} \mathcal{F}_i(\boldsymbol{w}_{o,l}^b, \zeta_{l-1}) \\
= ~&(\boldsymbol{w}_g - \boldsymbol{w}_{o,l}^a)\frac{\mu_{\beta}g_{o,l}^\mathrm{T}(\boldsymbol{w}_{o,l}^b)}{\sqrt{g}(1 + \mu_{\beta} \phi^i_{t_i-1})^{2}}(1-\frac{\upsilon^i_{t_i-1}}{\sqrt{g - o + 1}}).
\end{aligned}
\end{equation*}
Similarly, we can derive the partial deviation of the loss function with respect to $\upsilon^i_{t_i-1}$:
\begin{equation*}
\begin{aligned}
\nabla_{\upsilon^i_{t_i-1}} \mathcal{F}_i(\boldsymbol{w}_{o,l}^b, \zeta_{l-1}) &= \frac{\mu_{\beta}\gamma^i_{t_i-1}g_{o,l}^\mathrm{T}(\boldsymbol{w}_{o,l}^b)(\boldsymbol{w}_{o,l}^a - \boldsymbol{w}_g)}{\sqrt{g}\sqrt{g - o + 1}(1 + \mu_{\beta} \phi^i_{t_i-1})^{2}}.
\end{aligned}
\end{equation*}

Algorithm \ref{alg:device_passive} shows the device-side model update algorithm of \AName{}. First, an epoch number $l^*$ (time slot) for  requesting a fresh global model is set based on $\theta_{t-1}$ when $t = 1$ or $\mathcal{H}_{t_i-1}^i$ when $t\neq 1$ (Line \ref{alg:select}). In the $l^*$-th local epoch, the device sends a request to the server (Line \ref{alg:send}), and it waits for the fresh global model (Line \ref{alg:wait}). After receiving the fresh global model (Line \ref{alg:receiveFresh}), we exploit Equation \eqref{eq:betaUpdate} to update $\beta^i_{t_i-1}$ (Line \ref{alg:betaUpdate}), Equation \eqref{eq:deviceAggregation} to update $\boldsymbol{w}_{o,l-1}$ (Line \ref{alg:deviceAggregation}), Equation \eqref{eq:controlDevice} to update $\gamma^i_{t-1}$ and $\upsilon^i_{t-1}$ (Line \ref{alg:controlDeviceUpdate}), the reward values (Line \ref{alg:rewardUpdate}), $b_{t-1}$ with $\rho$ being a hyper-parameter (Line \ref{alg:btUpdate}), $\theta_{t}$ when $t = 1$ or $\mathcal{H}_{t_i-1}^i$ when $t \neq 1$ (Line \ref{alg:rlupdate}). Finally, the local model is updated (Line \ref{alg:SGDUpdate}).

\subsection{Convergence Analysis}

In this subsection, we present the assumptions of \AName{}, followed by its convergence guarantees.
\begin{assumption} 
\label{ass:smooth}
($L$-smoothness) The loss function $\mathcal{F}_i$ is differentiable and $L$-smooth for each device $i \in \mathcal{M}$ and $\forall x, y$, 
$
\mathcal{F}_i(y) - \mathcal{F}_i(x) \leq \ip{\nabla \mathcal{F}_i(x)}{y-x} + \frac{L}{2} \|y-x\|^2
$
with $L > 0$.
\end{assumption}
\begin{assumption} 
\label{ass:convex}
($\mu$-strongly convex)
The loss function $\mathcal{F}_i$ is $\mu$-strongly convex for each device $i \in \mathcal{M}$: $ \langle \nabla \mathcal{F}_i(x) - \nabla \mathcal{F}_i(y), x - y \rangle  \geq  \mu \parallel x - y \parallel ^ 2$ with $\mu > 0$.
\end{assumption}
\begin{assumption} 
\label{ass:sampling}
(Unbiased sampling)
The local sampling is unbiased and the local gradients are unbiased stochastic gradients $\mathbb{E}_{\zeta_l \sim \mathcal{D}_i} [\nabla \mathcal{F}_i(\boldsymbol{w}_{o,l}; \zeta_l)] = \nabla \mathcal{F}_i(\boldsymbol{w}_{o,l})$.
\end{assumption}
\begin{assumption} 
\label{ass:gradient}
(Bounded local gradient)
The stochastic gradients are bounded on each device $i \in \mathcal{M}$: $\mathbb{E}_{\zeta_l \sim \mathcal{D}_i} \parallel \nabla \mathcal{F}_i(\boldsymbol{w}_{o,l}; \zeta_l) \parallel^2 \le{\mathcal{G}^2}$.
\end{assumption}
\begin{assumption} 
\label{ass:variance}
(Bounded local variance)
The variance of local stochastic gradients are bounded on each device $i \in \mathcal{M}$ is bounded: $\mathbb{E}_{\zeta_l \sim \mathcal{D}_i} \parallel \nabla \mathcal{F}_i(\boldsymbol{w}_{o,l}; \zeta_l) - \mathcal{F}(\boldsymbol{w}_{o,l}) \parallel^2 \le{\mathcal{V}^2}$.
\end{assumption}
\begin{theorem}
Let Assumptions \ref{ass:smooth} - \ref{ass:variance} hold, after $T$ global updates, \AName{} converges to a critical point:
\begin{align*}
&\min_{t=0}^{T} \E\parallel \nabla \mathcal{F}(\boldsymbol{w}_{o,l}) \parallel^2 \\
\leq~& \frac{2 \E\left[ \mathcal{F}(\boldsymbol{w}_{0}) - \mathcal{F}(\boldsymbol{w}_{T}) \right]}{\alpha_{min} \mathcal{L}_{min}^3} + \OM\left( \frac{L \mathcal{G}^2 \mathcal{L}_{max}}{\mathcal{L}_{min}^3} \right) \\
&\quad + \OM\left( \frac{\mathcal{L}^i \mathcal{V}^2}{ \mathcal{L}_{min}^7}\right) + \OM\left( \frac{\tau \mathcal{G}^2 \mathcal{L}_{max}}{ \mathcal{L}_{min}^7} \right) \\
&\quad + \OM\left( \frac{ \mathcal{G}^2 \mathcal{L}_{max} }{\mathcal{L}_{min}^3} \right) + \OM\left( \frac{ L \mathcal{G}^2 \mathcal{L}_{max} }{\mathcal{L}_{min}^3} \right)\\
&  + \OM\left(\frac{ L \tau^2 \mathcal{G}^2 \mathcal{L}_{max}^2}{\mathcal{L}_{min}^3} \right) + \OM\left( \frac{ L \tau^2 \mathcal{G}^2 \mathcal{L}_{max}^2 }{\mathcal{L}_{min}^3} \right),
\end{align*}
where $\alpha_{min} \leq \alpha_t^i$, $\mathcal{L}_{min} \leq \mathcal{L}_t \leq \mathcal{L}_{max}$, $\eta_i = \frac{1}{\sqrt{T}}$, $\forall i \in \mathcal{M}$, and $T = \mathcal{L}_{min}^6$.
\end{theorem}

\rev{
See Appendix \ref{sec:appendx} for the detailed proof of Theorem 1, which demonstrates how Assumptions \ref{ass:smooth} - \ref{ass:variance} ensure convergence.
}

\section{Synchronous Federated Learning with Timely Update Dissemination}
\label{sec:sync}

\begin{figure}

\begin{algorithm}[H]
\caption{\rev{\SName{} on the Server and Devices}}
\label{alg:FedSSMU}
\begin{algorithmic}[1]

\REQUIRE \quad \\
\textbf{Server Inputs:} \\
$T$: The maximum number of global rounds \\
$m'$: The number of devices to be triggered in each round \\
$\tau$: The predefined staleness limit \\
$\boldsymbol{w}_{1,0}$: The initial global model for round 1 \\
$\lambda_{0}^i, \sigma_{0}^i, \iota_{0}^i$: Initial control parameters for each device $i$ \\
$\eta_{\lambda^i}, \eta_{\sigma^i}, \eta_{\iota^i}$: Learning rates for server-side control parameter updates \\

\textbf{Device Inputs:} \\
$\mathcal{L}^i$: The maximum number of local epochs on Device $i$ \\
$\boldsymbol{w}_{g,0}$: The global model received from the server at round $g$ \\
$\gamma_{t_i}^i, \upsilon_{t_i}^i$: Initial control parameters for device-side aggregation \\
$\eta_{\gamma^i}, \eta_{\upsilon^i}$: Learning rates for device-side control parameter updates \\
$\eta_i$: Learning rate for local SGD \\
$\Theta_g$: Parameters of the RL model at round $g$ \\

\ENSURE \quad \\
$\boldsymbol{w}_{T+1,0}$: The final global model after round $T$ \\

\STATE \textbf{Server-Side Procedure:}
\FOR{$g = 1$ to $T$}
    \STATE Randomly select $m'$ devices and send $\boldsymbol{w}_{g,0}$ to them \quad // Trigger devices with global model
    \STATE Initialize $k = 0$ \quad // Counter for aggregated models in this round
    \WHILE{not all $m'$ devices have uploaded their models}
        \STATE Receive $\boldsymbol{w}_{g,0}^i$ from Device $i$ \quad // Collect local model from device
        \STATE Compute staleness $\tau_i = k + 1$ \quad // Staleness based on aggregation count
        \IF{$\tau_i > \tau$}
            \STATE Discard $\boldsymbol{w}_{g,0}^i$ \quad // Discard if staleness exceeds limit
        \ELSE
            \STATE Update $\lambda_g^i, \sigma_g^i, \iota_g^i$ using Equation \eqref{eq:control} \quad // Dynamic parameter adjustment
            \STATE Compute $\alpha_g^i$ using Equation \eqref{eq:alpha} \quad // Calculate aggregation weight
            \STATE Update $\boldsymbol{w}_{g,k+1} = (1 - \alpha_g^i) \boldsymbol{w}_{g,k} + \alpha_g^i \boldsymbol{w}_{g,0}^i$ \quad // Aggregate models
            \STATE Set $k = k + 1$ \quad // Increment aggregation counter
        \ENDIF
    \ENDWHILE
    \STATE Set $\boldsymbol{w}_{g+1,0} = \boldsymbol{w}_{g,k}$ \quad // Set global model for next round
\ENDFOR
\STATE \textbf{Return} $\boldsymbol{w}_{T+1,0}$

\STATE \textbf{Device-Side Procedure (for each Device $i$ at round $g$):}
\STATE Receive $\boldsymbol{w}_{g,0}$ from server and set $\boldsymbol{w}_{g,0}^i = \boldsymbol{w}_{g,0}$
\STATE Determine request time slot $l^*$ using RL model with parameters $\Theta_g$ \quad // See Section \ref{subsec:slot_selection}
\FOR{$l = 1$ to $\mathcal{L}^i$}
    \IF{$l = l^*$}
        \STATE Request fresh global model from server
        \STATE Receive $\boldsymbol{w}_{g,k}$ (if available, where $k > 0$)
    \ENDIF
    \IF{$\boldsymbol{w}_{g,k}$ received and $k > 0$}
        \STATE Compute $\beta_{t_i}^i$ using Equation \eqref{eq:betaUpdate} \quad // Weight for merging
        \STATE Update $\boldsymbol{w}_{g,l-1}^i = (1 - \beta_{t_i}^i) \boldsymbol{w}_{g,l-1}^i + \beta_{t_i}^i \boldsymbol{w}_{g,k}$ \quad // Merge with fresh global model
        \STATE Update $\gamma_{t_i}^i, \upsilon_{t_i}^i$ using Equation \eqref{eq:controlDevice} \quad // Adjust parameters
    \ENDIF
    \STATE Update $\boldsymbol{w}_{g,l}^i = \boldsymbol{w}_{g,l-1}^i - \eta_i \nabla \mathcal{F}_i(\boldsymbol{w}_{g,l-1}^i, \zeta_{l-1})$ \quad // Local SGD, $\zeta_{l-1} \sim \mathcal{D}_i$
\ENDFOR
\STATE Upload $\boldsymbol{w}_{g,\mathcal{L}^i}^i$ to server

\end{algorithmic}
\end{algorithm}
\end{figure}

In this section, we extend the above approach to the synchronous FL case. In the synchronous version named FedSSMU as shown in Algorithm \ref{alg:FedSSMU}, the server triggers local training on $m'$ devices at the beginning of each global round. The training is organized into multiple global rounds, starting with an initial global model version of 0. The global model version is incremented by 1 at the end of each global round. 

The training process of \SName{} includes the following steps:
\begin{itemize}
\item Step \textcircled{1}: At the beginning of each global round $g$, the server triggers $m' (m' \leq m)$ randomly selected devices and broadcasts the current global model $\boldsymbol{w}_{g,0}$ to them, where $g$ is the global round number, and $0$ indicates that no local models have been received yet by the server in this round.
\item Step \textcircled{2}: Each selected device performs local training using its local dataset. The training is conducted by applying the gradient descent method described in Equation \eqref{eq:localSGD}.
\item 
Step \textcircled{3}: During local training, Device $i$ dynamically determines the slot for requesting the latest global model $\boldsymbol{w}_{g, k}$ from the server to minimize staleness, based on the RL-based method detailed in Section \ref{subsec:slot_selection}. This mechanism balances the trade-off between requesting the updated global model too early, which risks incorporating insufficient updates, and requesting too late, which may lead to underutilizing the updated model.
\item Step \textcircled{4}: When the server receives a request for an updated model, it checks the version of the global model. If the version $k > 0$, indicating that there is a newer global model, the server sends the updated global model $\boldsymbol{w}_{g,k}$ to the requesting device.
\item Step \textcircled{5}: Upon receiving the updated global model, the device merges it with its latest local model using the aggregation method outlined in Section \ref{subsec:device_agg}. This process is governed by Equation \eqref{eq:deviceAggregation} and includes parameter adjustments as specified in Equation \eqref{eq:controlDevice}. The device then continues local training with this updated model.
\item Step \textcircled{6}: Once local training is completed, Device $i$ uploads its locally trained model to the server.
\item Step \textcircled{7}: Whenever the server receives a local model $\boldsymbol{w}^i_{g,0}$ from any device, it immediately aggregates this local model with the current global model $\boldsymbol{w}_{g,k}$ using the server-side aggregation method described in Section \ref{subsec:MAServer}. This process adheres to Equation \eqref{eq:serverAggregation} and incorporates parameter adjustments as defined in Equation \eqref{eq:control}. After each aggregation, the global model version is incremented from $\boldsymbol{w}_{g,k}$ to $\boldsymbol{w}_{g,k+1}$, where $k$ denotes the number of local models that have been aggregated so far in the current global round $g$. The staleness of the local model is calculated as $\tau_i = k + 1$. If $\tau_i$ exceeds a predefined threshold $\tau$, indicating excessive staleness, the local model is discarded to maintain convergence and avoid reverting the global model to a less accurate state.
\end{itemize}

By following this process, the synchronous FL framework ensures efficient and up-to-date global model training while maintaining convergence within acceptable staleness bounds. The convergence property in Theorem 1 also applies to the synchronous case, since we can regard the synchronous case as a special asynchronous one, where the server now triggers $m'$ devices at the start of each global round and waits until receiving $m'$ local models.

\begin{table}[t]
\centering
\caption{The network structure of CNN.}
\label{tbl:cnn}
\resizebox{0.6\hsize}{!}{
\begin{tabular}{|l|l|}
\hline 
Layer (type) & Parameters \\ \hline 
conv1(Convolution)& channels=64, kernel\_size=2 \\ \hline 
activation1(Activation)& null \\ \hline 
conv2(Convolution)& channels=32, kernel\_size=2 \\ \hline 
activation2(Activation)& null \\ \hline 
flatten1(Flatten)& null \\ \hline 
dense1(Dense)& units=10 \\ \hline 
softmax(SoftmaxOutput)& null \\ \hline 
\end{tabular} 
}
\end{table}

\begin{table*}[!t]
\caption{Values of hyper-parameters in the experimentation.}
\label{tab:parameters1}
\begin{center}
\resizebox{\hsize}{!}{
\begin{tabular}{c|c|c|c|c|c|c|c|c|c|c|c|c}
\toprule
\multicolumn{1}{c|}{\multirow{3}{*}{Name}} & \multicolumn{12}{c}{Values} \\
\cline{2-13}
\multicolumn{1}{c|}{}& \multicolumn{3}{c|}{LeNet} & \multicolumn{2}{c|}{CNN} & \multicolumn{2}{c|}{ResNet} & \multicolumn{2}{c|}{AlexNet} & \multicolumn{2}{c|}{VGG} & \multicolumn{1}{c}{TextCNN} \\
\cline{2-13}
\multicolumn{1}{c|}{} & FMNIST & CIFAR-10 & CIFAR-100  & CIFAR-10 & CIFAR-100 & CIFAR-100 & Tiny-ImageNet & CIFAR-10 & CIFAR-100 & CIFAR-10 & CIFAR-100 & IMDb  \\
\hline
$m$ & 100 & 100 & 100 & 100 & 100 & 100 & 100 & 100 & 100 & 100 & 100 & 100  \\
$m'$ & 10 & 10 & 10 & 10 & 10 & 10 & 10  & 10 & 10 & 10 & 10 & 10  \\
$T$ & 500 & 500 & 500 & 500 & 500 & 500 & 500 & 500 & 500 & 500 & 500 & 500 \\
$\tau$ & 99 & 99 & 99 & 99 & 99 & 99 & 99 & 99 & 99 & 99 & 99 &99\\
$\mathcal{T}$ & 10 & 10 & 10 & 10 & 10 & 10 & 10  & 10 & 10 & 10 & 10 & 10 \\
$\eta_{\lambda^i}$ &0.0001	&0.001	&0.0001	&0.001	&0.00001	&0.0001	&0.0001  &0.0001	&0.0001	&0.0001	&0.0001	&0.0001 \\
$\eta_{\sigma^i}$ &0.0001	&0.001	&0.0001	&0.001	&0.00001	&0.0001	&0.0001 &0.0001	&0.0001	&0.0001	&0.0001	&0.0001 \\
$\eta_{\iota^i}$ &0.0001	&0.1	&0.0001	&0.0001	&0.00001	&0.0001	&0.0001 &0.0001	&0.0001	&0.0001	&0.0001	&0.0001 \\
$\eta_{\gamma^i}$ &0.0001	&0.1	&0.0001	&0.1	&0.00001	&0.0001	&0.0001 &0.0001	&0.0001	&0.0001	&0.0001	&0.0001  \\
$\eta_{\upsilon^i}$&0.0001	&0.001	&0.0001	&0.001	&0.00001	&0.0001	&0.0001 &0.0001	&0.0001	&0.0001	&0.0001	&0.0001  \\
$\eta_i$ &0.005	&0.03	&0.03	&0.028	&0.013	&0.03	&0.03   &0.03	&0.03	&0.03	&0.03	&0.001 \\
$\eta^i_{RL}$ & 0.001 & 0.001 & 0.001 & 0.001 & 0.001 & 0.001 & 0.001 & 0.001 & 0.001 & 0.001 & 0.001 & 0.001 \\
\bottomrule
\end{tabular}
}
\end{center}
\end{table*}

\section{Experiments}
\label{sec:exp}

In this section, we present a comparative experimental analysis of \AName{} and \SName{} against 9 state-of-the-art methods. We begin by outlining the experimental setup, followed by a detailed presentation of the results.

\subsection{Experimental Setup}
\label{sec:evaluation_setup}

\subsubsection{Baselines}
\rev{To comprehensively evaluate the performance of the proposed method, we compared it with a series of asynchronous and synchronous federated learning baselines. Asynchronous baselines include: FedAsync \cite{xie2019asynchronous}, which handles asynchronicity by weighting stale updates; PORT \cite{Su2022How}, which reportedly uses polynomial functions to evaluate update latency and divergence; ASO-Fed \cite{chen2020asynchronous}, which adopts an attention-like mechanism on the server side to handle non-IID data; FedBuff \cite{Nguyen2022FedBuff}, which buffers multiple client updates before aggregation; and FedSA \cite{chen2021fedsa}, a staleness-aware method for non-IID data. Synchronous baselines include: the classic FedAvg \cite{mcmahan2017communication}; FedProx \cite{Li2020FedProx}, which addresses data heterogeneity through a proximal term; MOON \cite{li2021model}, which leverages model contrastive learning to mitigate heterogeneity; FedDyn \cite{acar2021federated}, which uses dynamic regularization to align local and global objectives; and FedLWS \cite{shi2025fedlws}, which employs adaptive layer-wise weight shrinking to enhance model generalization by dynamically computing shrinking factors for each layer without requiring a proxy dataset. These baselines collectively represent mainstream approaches in the federated learning field for addressing efficiency and data heterogeneity challenges.}

\subsubsection{Datasets}
We utilize 5 public datasets, i.e., Fashion-MNIST (FMNSIT) \cite{xiao2017fashion}, CIFAR-10 and CIFAR-100 \cite{krizhevsky2009learning}, IMDb \cite{zhou2021distilled}, and Tiny-ImageNet \cite{le2015tiny}. The data on each device is non-IID based on a Dirichlet distribution \cite{li2021federated}. 

\subsubsection{Models}
We leverage 6 models to deal with the data, i.e., LeNet5 (LeNet) \cite{lecun1989handwritten}, a synthetic CNN network (CNN) with the network structure shown in Table~\ref{tbl:cnn}, ResNet20 (ResNet) \cite{He2016}, AlexNet \cite{Krizhevsky2012ImageNet}, TextCNN \cite{zhou2021distilled}, and VGG-11 (VGG) \cite{simonyan2015very}.

\subsubsection{Environment}

We utilize 44 Tesla V100 GPUs to create a simulation of the FL environment. The setup includes a server and 100 heterogeneous devices. Device heterogeneity is modeled by varying local training durations; specifically, the slowest device's training time is five times that of the fastest device, with each device's training time randomly sampled within this range. We apply a learning rate decay during the training process. We meticulously fine-tune the hyper-parameters for each method, reporting the optimal settings in this paper. The hyper-parameter values are detailed in Table \ref{tab:parameters1}.

\rev{
\subsubsection{Performance Metrics}
We evaluate the performance of FedASMU and FedSSMU using two key metrics: convergence accuracy and training time to reach a target accuracy. Convergence accuracy measures the final performance of the global model, reflecting its ability to generalize across heterogeneous data. Training time assesses the efficiency of the FL process, crucial in heterogeneous environments where device capabilities vary. These metrics were selected to comprehensively capture both the effectiveness and efficiency of our proposed methods, aligning with the goals of addressing statistical and system heterogeneity.
}

\subsection{Evaluation}
\label{sec:expResults}

\begin{table*}[!t]
  \caption{The accuracy and training time with \AName{}, \SName{} and diverse baseline approaches. ``Acc'' is the convergence accuracy of the global model. ``Time'' refers to the training time to achieve a target accuracy, i.e., 0.40 for AlexNet with CIFAR-10,  0.12 for AlexNet with CIFAR-100, 0.45 for VGG with CIFAR-10, 0.12 for VGG with CIFAR-100, 0.85 for TextCNN, and 0.70 for LeNet with FMNIST, 0.30 for LeNet with CIFAR-10,  0.13 for LeNet with CIFAR-100, 0.40 for CNN with CIFAR-10, 0.15 for CNN with CIFAR-100, 0.25 for ResNet with CIFAR-100, and 0.12 for ResNet with Tiny-ImageNet. ``/'' represents that the method does not achieve the target accuracy.
  }
  \label{tab:cmp_overall}
  \centering

  \resizebox{\hsize}{!}{
  \begin{tabular}{l|lr|lr|lr|lr|lr|lr}
    \toprule
    \multirow{3}{*}{Method} & \multicolumn{4}{c|}{AlexNet}  & \multicolumn{4}{c|}{VGG}  & \multicolumn{2}{c|}{TextCNN} & \multicolumn{2}{c}{LeNet} \\
    \cmidrule(r){2-13} & \multicolumn{2}{c|}{ CIFAR-10}  & \multicolumn{2}{c|}{ CIFAR-100} & \multicolumn{2}{c|}{ CIFAR-10}  & \multicolumn{2}{c|}{ CIFAR-100}  & \multicolumn{2}{c|}{IMDb} & \multicolumn{2}{c}{FMNIST} \\
    \cmidrule(r){2-13} & Acc  & Time & Acc  & Time & Acc & Time & Acc & Time & Acc & Time & Acc & Time \\
    \midrule
    \AName{} (Our)  &  \textbf{0.490}     & \textbf{12591}   &  \textbf{0.246}    & \textbf{12150} &  \textbf{0.657}    & \textbf{47356}   &  \textbf{0.267}    & \textbf{83226}    &  \textbf{0.882}   &  \textbf{3994}      &  \textbf{0.858}   &  \textbf{11603} \\
    FedAsync \rev{\cite{xie2019asynchronous}}   & 0.415  & 83693      & 0.204  & 13717          & 0.645  & 77530    & 0.168    & 375236      & 0.875  & 5837        & 0.839  & 15941         \\
    PORT    \rev{\cite{Su2022How}}    & 0.370  & /    & 0.193  & 17400               & 0.561  & 97848        & 0.213   & 120533          & 0.876  & 4884     & 0.799  & 75323         \\
    ASO-Fed  \rev{\cite{chen2020asynchronous}}   & 0.450  & 128015     & 0.243  & 60864          & 0.548  & 276352   & 0.131   & 405906     & 0.811  & /       & 0.833  & 42472        \\
    FedBuff  \rev{\cite{Nguyen2022FedBuff}}   & 0.471  & 27763      & 0.226  & 27672           & 0.617  & 192419        & 0.258   & 167053      & 0.876  & 7671     & 0.827  & 16953        \\
    FedSA   \rev{\cite{chen2021fedsa}}    & 0.410  & 22198       & 0.173 & 17508               & 0.417  & /    & 0.187   & 120426      & 0.865  & 5251        & 0.786  & 14473            \\
    \midrule
        \SName{} (Our)  &  \textbf{0.496}     & \textbf{56655}   &  \textbf{0.255}    & \textbf{34127} &  0.612    & 228078   &  \textbf{0.268}    & \textbf{243896}    &  \textbf{0.883}   &  {16940}      &  \textbf{0.858}   &  \textbf{13753} \\
    FedAvg   \rev{\cite{mcmahan2017communication}}   & 0.446   & 157678     & 0.204  & 92558           & 0.519  & 346890    & 0.115   & /     & 0.874 &  13960   & 0.780  & 97692        \\
    FedProx \rev{\cite{Li2020FedProx}}    & 0.446  & 141125    & 0.204  & 91369          & 0.522  & 346890        & 0.109   & /      & 0.875  & 15668    & 0.778  & 97692         \\
    MOON   \rev{\cite{li2021model}}     & 0.440  & 157678    & 0.197  & 89297           & 0.522   & 346890   & 0.112   & /    & 0.875  & 13960   & 0.785  & 97692       \\
    FedDyn   \rev{\cite{acar2021federated}}   & 0.439  & 144999         & 0.193  & 103950           & 0.549  & 195032    & 0.224   & 307955      & 0.874  & \textbf{12674}   & 0.842  & 51223             \\
    \rev{FedLWS   \cite{shi2025fedlws}}   & \rev{0.458}  & \rev{125514}         & \rev{0.226}  & \rev{58802}           & \rev{\textbf{0.643}}  & \rev{\textbf{180782}}    & \rev{0.177}   & \rev{330701}      & \rev{0.874}  & \rev{17988}   & \rev{0.846}  & \rev{20599}             \\
    \bottomrule
  \end{tabular}
  }
    \resizebox{\hsize}{!}{
  \begin{tabular}{l|lr|lr|lr|lr|lr|lr}
    \toprule
    \multirow{3}{*}{Method} & \multicolumn{4}{c|}{LeNet}  & \multicolumn{4}{c|}{CNN}  & \multicolumn{4}{c}{ResNet} \\
    \cmidrule(r){2-13} & \multicolumn{2}{c|}{ CIFAR-10}  & \multicolumn{2}{c|}{ CIFAR-100} & \multicolumn{2}{c|}{ CIFAR-10}  & \multicolumn{2}{c|}{ CIFAR-100}  & \multicolumn{2}{c|}{ CIFAR-100}  & \multicolumn{2}{c}{Tiny-ImageNet}\\
    \cmidrule(r){2-13} & Acc  & Time & Acc  & Time & Acc & Time & Acc & Time & Acc & Time & Acc & Time\\
    \midrule
    \AName{} (Our)  &  \textbf{0.485}    & \textbf{8800}   &  \textbf{0.183}    & \textbf{20737} &  \textbf{0.603}    & \textbf{10109}   &  \textbf{0.279}    & \textbf{30569}    &  \textbf{0.359}   &  \textbf{16027}    &  \textbf{0.168}   &  \textbf{22415}       \\
    FedAsync    & 0.483  & 36565      & 0.168  & 102113          & 0.491  & 24931    & 0.230    & 37160      & 0.316  & 21107       & 0.143   & 31288          \\
    PORT        & 0.307  & 365642    & 0.106  & /               & 0.380  & /        & 0.146   & /          & 0.314  & 35712       & 0.133   & 78155     \\
    ASO-Fed     & 0.411  & 83973     & 0.151  & 110942          & 0.477  & 92246    & 0.209   & 103090     & 0.280  & 198797      & 0.125   & 359899         \\
    FedBuff     & 0.366  & 9907      & 0.174  & 25791           & 0.366  & /        & 0.202   & 65736      & 0.317  & 27672       & 0.148   & 43523      \\
    FedSA       & 0.306  & 21562     & 0.0836 & /               & 0.509  & 20415    & 0.191   & 94169      & 0.195  & /           & 0.116  & /           \\
    \midrule
    \SName{} (Our)  &  \textbf{0.496}    & \textbf{63500}   &  \textbf{0.220}    & {98303} &  \textbf{0.628}    & {44338}   &  \textbf{0.293}    & {104844}    &  \textbf{0.378}   &  \textbf{35145}    &  \textbf{0.193}   &  \textbf{69848}       \\
    FedAvg      & 0.450   & 125514     & 0.167  & 95306           & 0.562  & 117794    & 0.249   & 73145      & 0.298  & 109680      & 0.145   & 155023     \\
    FedProx     & 0.374  & 126958    & 0.173  & {93430}           & 0.360  & /        & 0.244   & 73145      & 0.303  & 109680      & 0.149   & 151935        \\
    MOON        & 0.305  & 437531    & 0.170  & {93430}           & 0.478   & 100302   & 0.225   & 252703     & 0.304  & 106021      & 0.151   & 139444    \\
    FedDyn      & 0.277  & /         & 0.150  & \textbf{70260}           & 0.508  & \textbf{43974}    & 0.189   & \textbf{52874}      & 0.327  & 73711       & 0.139   & 103661          \\
    \rev{FedLWS}   & \rev{0.488}  & \rev{56019}         & \rev{0.181}  & \rev{91686}           & \rev{0.570}  & \rev{56019}    & \rev{0.264}   & \rev{108906}      & \rev{0.352}  & \rev{62891}   & \rev{0.176}  & \rev{92180}             \\
    \bottomrule
  \end{tabular}
  }
\end{table*}

\begin{figure*}[!t]
\centering
\subfigure[AlexNet \& CIFAR-10]{
\includegraphics[width=0.31\linewidth]{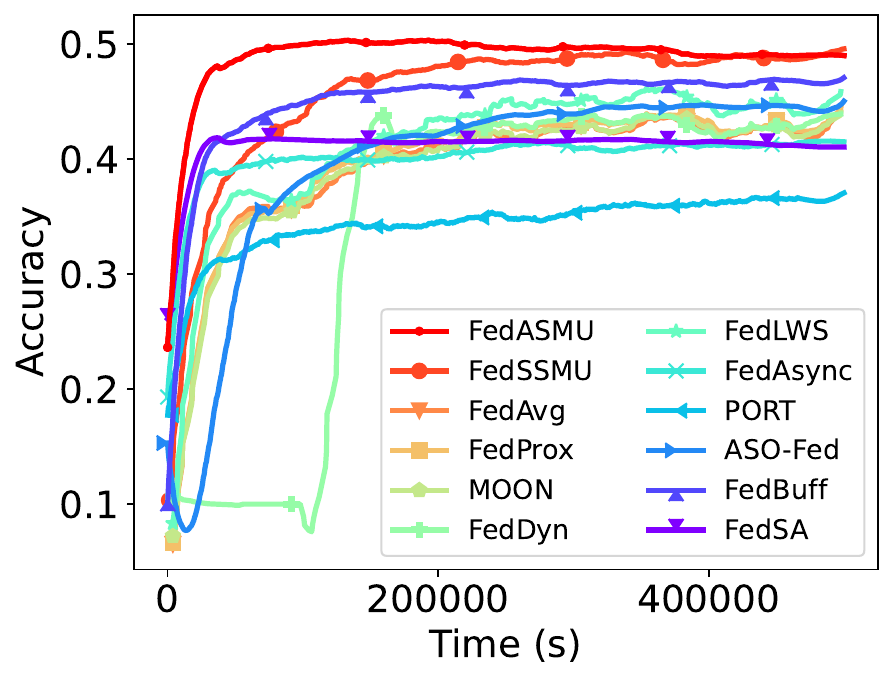}
\label{fig:cmp_dh_lenet_10}
}
\subfigure[AlexNet \& CIFAR-100]{
\includegraphics[width=0.31\linewidth]{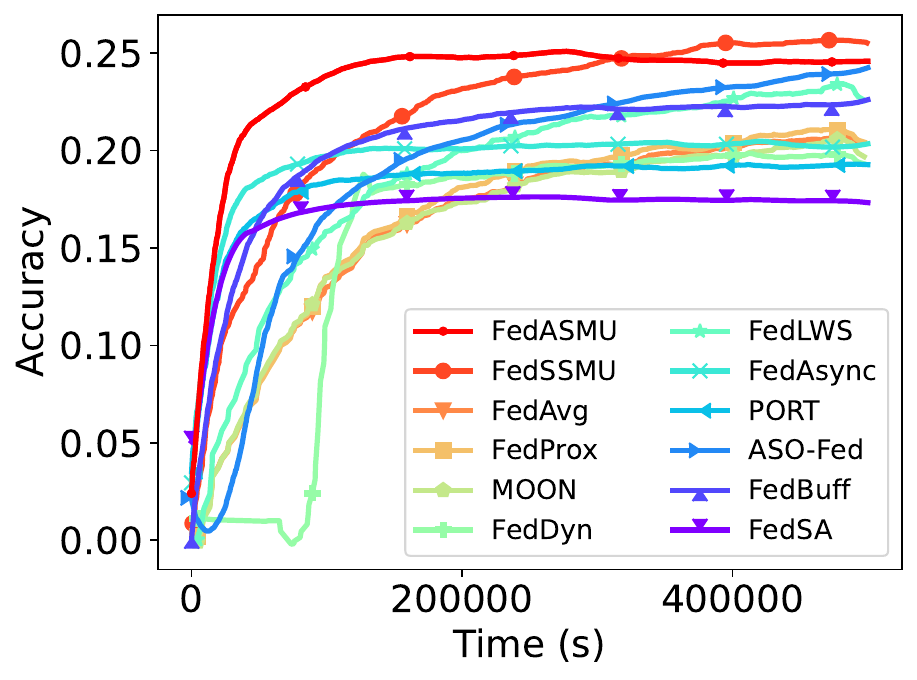}
\label{fig:cmp_dh_cnn_10}
}
\subfigure[VGG \& CIFAR-10]{
\includegraphics[width=0.31\linewidth]{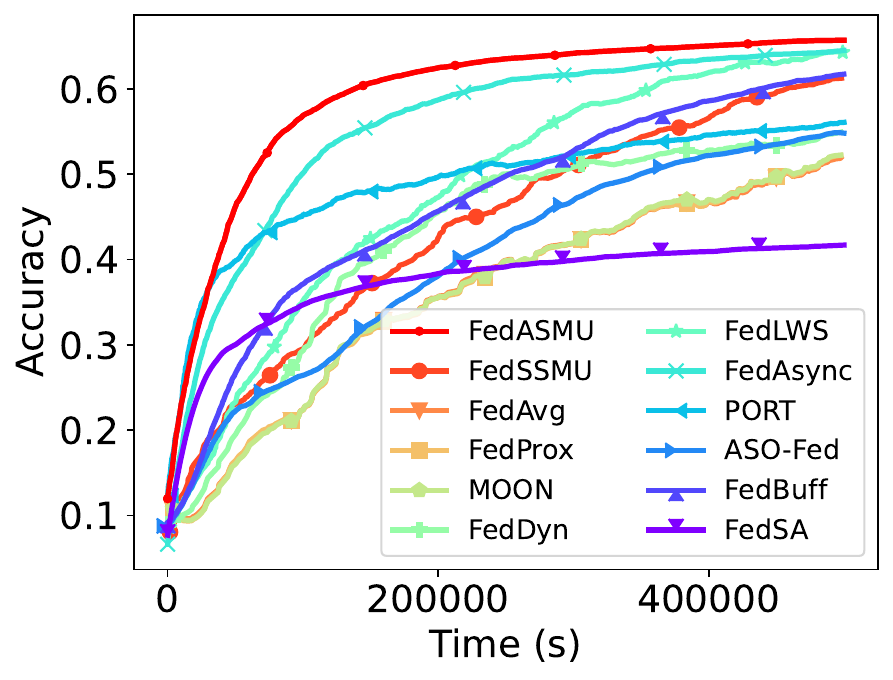}
\label{fig:cmp_dh_resnet_tiny}
}
\caption{The accuracy and training time with AlexNet \& CIFAR-10, AlexNet \& CIFAR-100, VGG \& CIFAR-10.}
\label{fig:async_CNN_CIFAR10}
\end{figure*}

\begin{figure*}[!t]
\centering
\subfigure[VGG \& CIFAR-100]{
\includegraphics[width=0.31\linewidth]{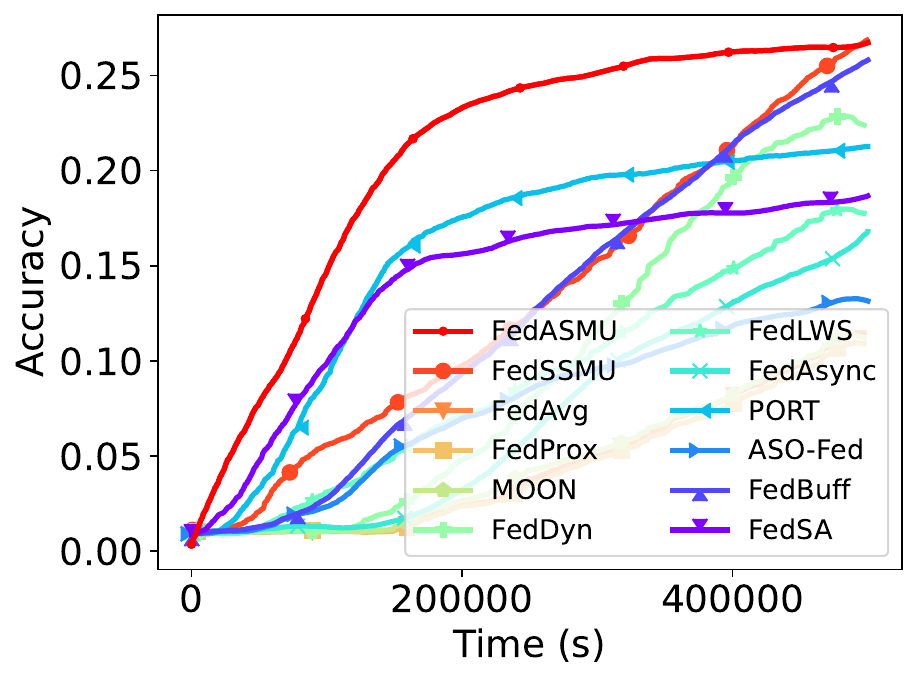}
\label{fig:cmp_dh_lenet_cifar100}
}
\subfigure[TextCNN \& IMDb]{
\includegraphics[width=0.31\linewidth]{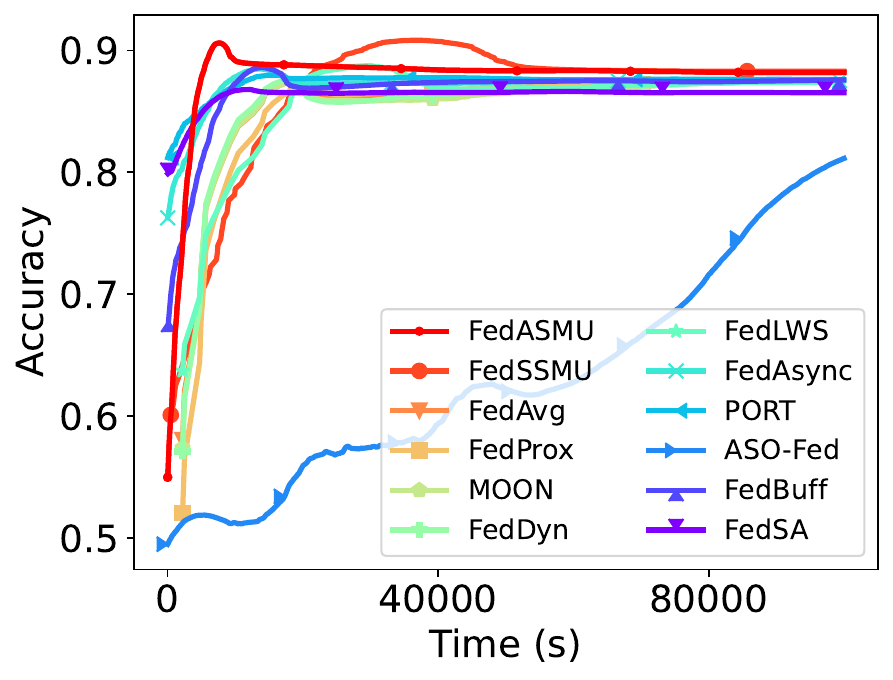}
\label{fig:cmp_dh_cnn_cifar100}
}
\subfigure[LeNet \& FMNIST]{
\includegraphics[width=0.31\linewidth]{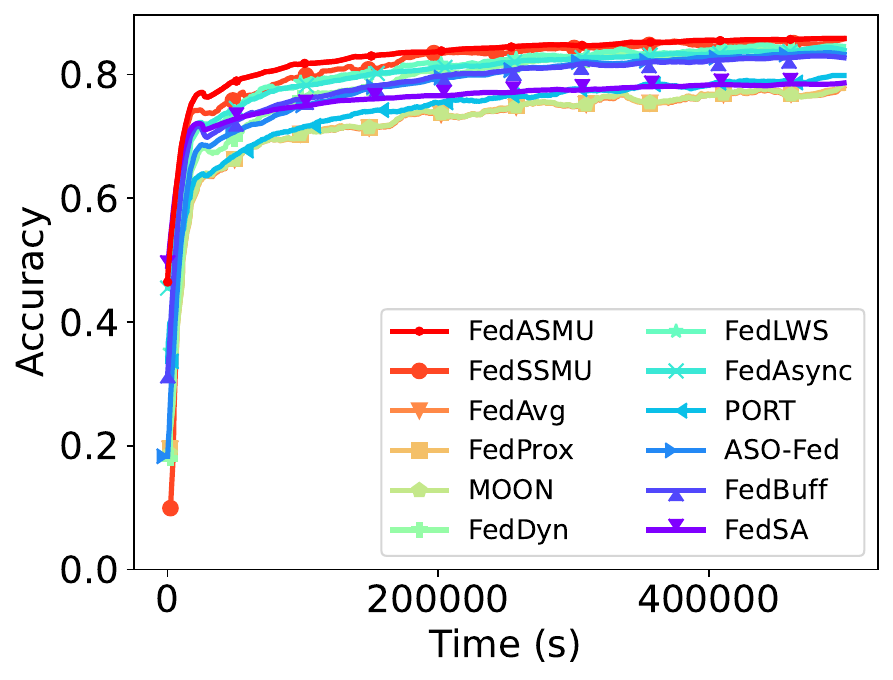}
\label{fig:cmp_dh_resnet_cifar100}
}
\caption{The accuracy and training time with VGG \& CIFAR-100, TextCNN \& IMDb, LeNet \& FMNIST.}
\label{fig:async_cmp}
\end{figure*}

\begin{figure*}[!htbp]
\centering
\subfigure[LeNet \& CIFAR-10]{
\includegraphics[width=0.31\linewidth]{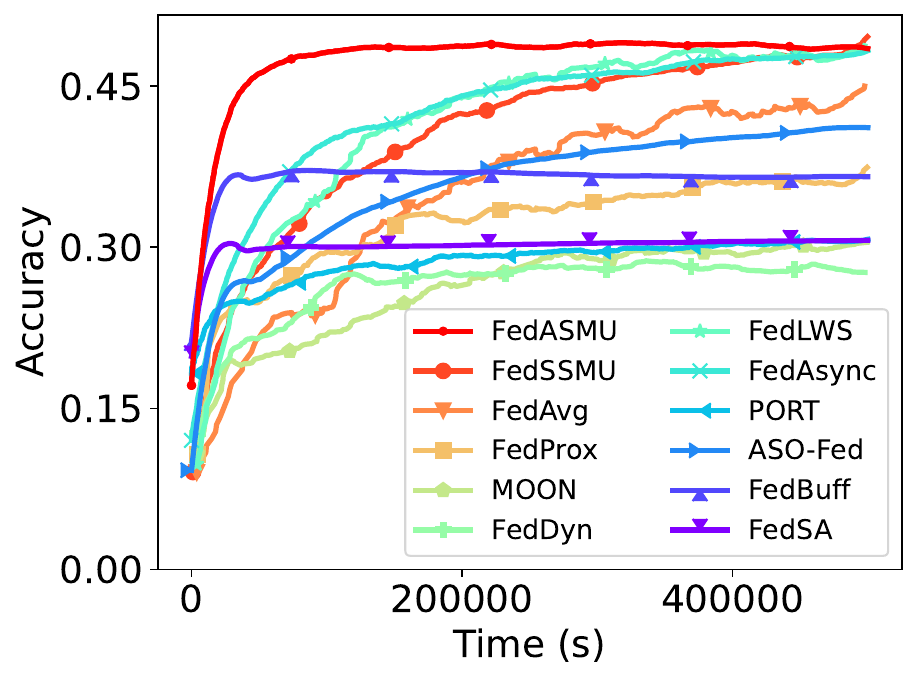}
\label{fig:alexnet_CIFAR10_cmp}
}
\subfigure[LeNet \& CIFAR-100]{
\includegraphics[width=0.31\linewidth]{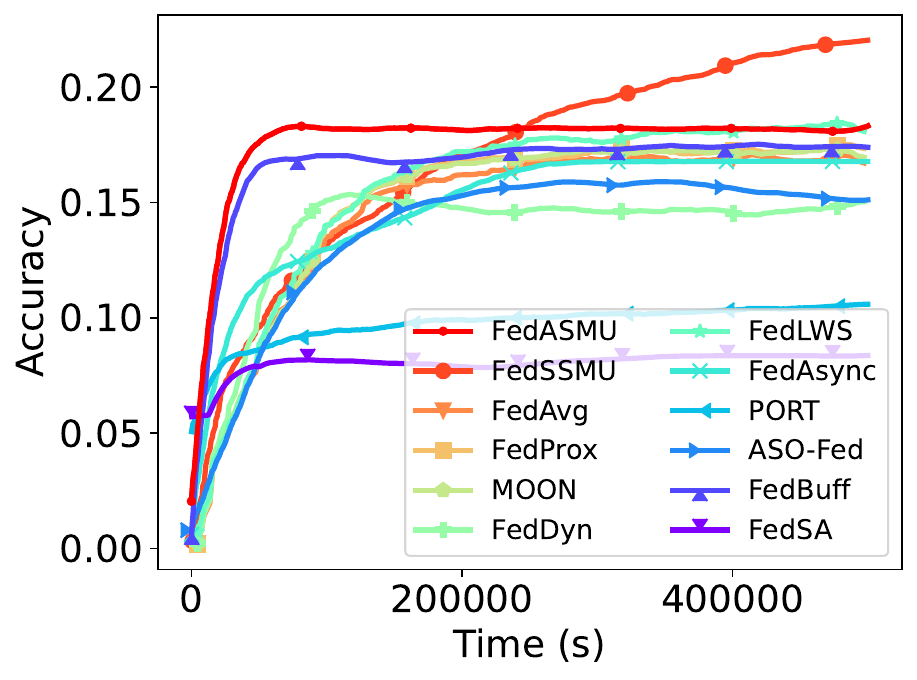}
\label{fig:vgg11_CIFAR10_cmp}
}
\subfigure[CNN \& CIFAR-10]{
\includegraphics[width=0.31\linewidth]{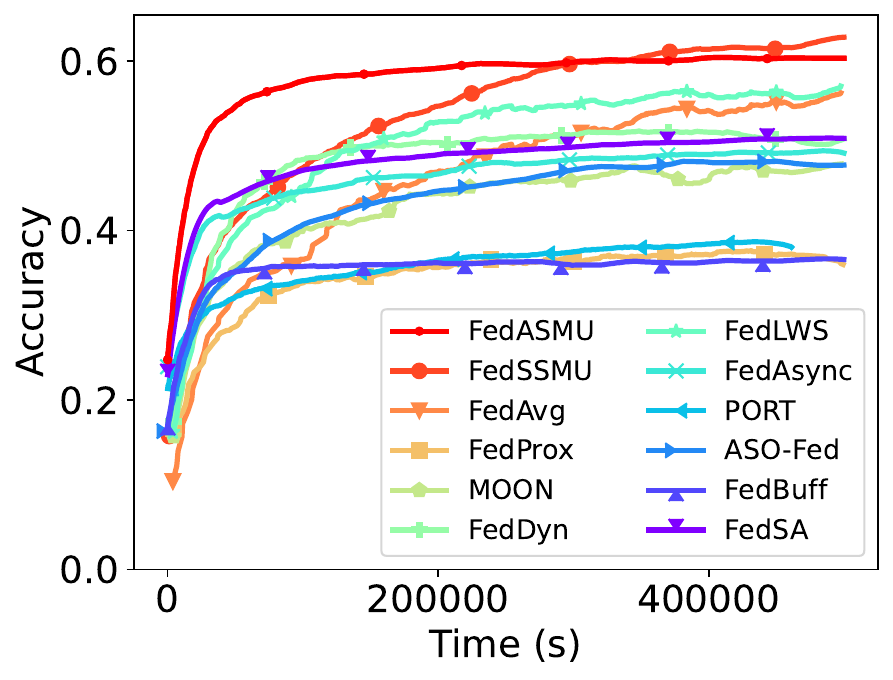}
\label{fig:textcnn_cmp}
}
\caption{The accuracy and training time with LeNet \& CIFAR-10, LeNet \& CIFAR-100, CNN \& CIFAR-10.}
\label{fig:async_cmp_cifar-10-IMDb}
\end{figure*}

\begin{figure*}[!htbp]
\centering
\subfigure[CNN \& CIFAR-100]{
\includegraphics[width=0.31\linewidth]{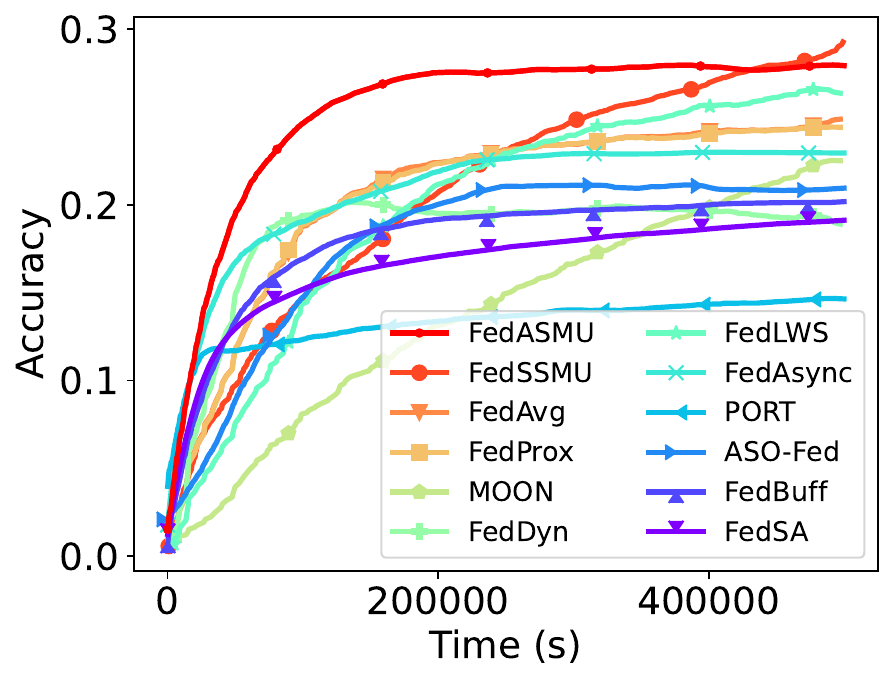}
\label{fig:alexnet_CIFAR100_cmp}
}
\subfigure[ResNet \& CIFAR-100]{
\includegraphics[width=0.31\linewidth]{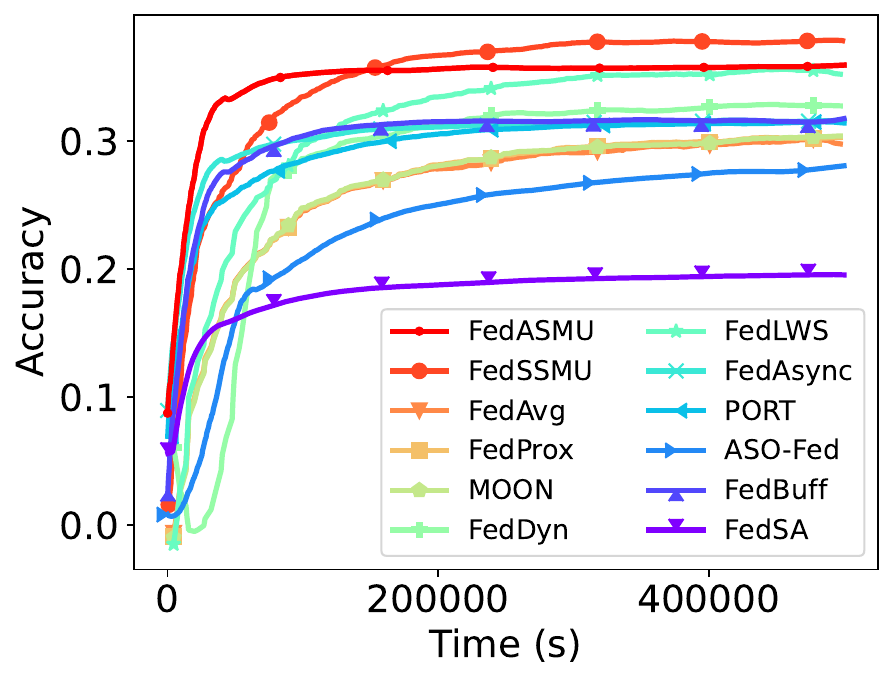}
\label{fig:vgg11_CIFAR100_cmp}
}
\subfigure[ResNet \& Tiny-ImageNet]{
\includegraphics[width=0.31\linewidth]{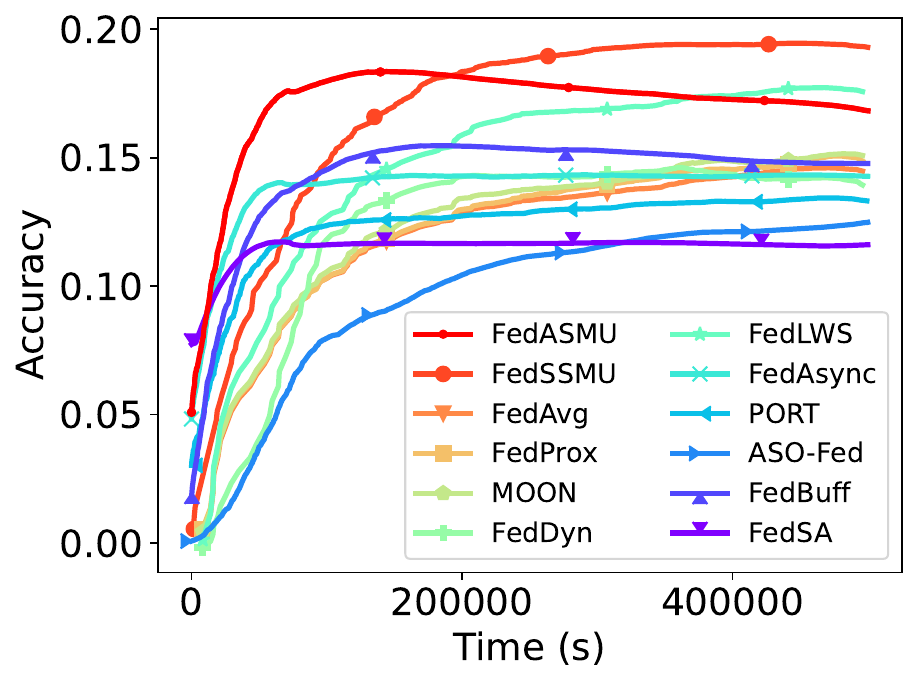}
\label{fig:lenet_FMNIST_cmp}
}
\caption{The accuracy and training time with CNN \& CIFAR-100, ResNet \& CIFAR-100, ResNet \& Tiny-ImageNet.}
\label{fig:async_cmp_cifar-100-FMNIST}
\end{figure*}

\subsubsection{Performance Overview}

As shown in Table \ref{tab:cmp_overall}, \AName{} and \SName{} consistently correspond to the highest convergence accuracy and training speed. Compared to asynchronous baseline methods, \AName{} reaches the target accuracy with significant speed improvements {(achieving gains from 11.42\% to 84.96\% over FedAsync, 18.22\% to 97.59\% over PORT, 70.35\% to 93.77\% over ASO-Fed, 11.17\% to 75.39\% over FedBuff, and 19.83\% to 67.54\% over FedSA)}. Furthermore, \AName{} achieves considerably higher accuracy {(outperforming FedAsync by 0.41\% to 58.93\%, PORT by 0.68\% to 91.10\%, ASO-Fed by 1.23\% to 103.82\%, FedBuff by 0.68\% to 64.75\%, and FedSA by 1.97\% to 118.90\%)}.

Compared with synchronous baseline methods, \SName{} demonstrates a substantially higher training speed, surpassing {FedAvg, FedProx and Moon by 34.25\% to 85.92\%, FedDyn by 20.80\% to 73.15\% and FedLWS by 3.73\% to 54.86\%}. Additionally, \SName{} achieves better convergence accuracy than these baselines, with gains of {1.03\% to 133.04\% over FedAvg, 0.91\% to 145.87\% over FedProx, 0.91\% to 139.29\% over MOON, 1.03\% to 79.06\% over FedDyn and 1.03\% to 51.41\% over FedLWS}. 

The experimental results, visualized with various baseline approaches and under normal bandwidth conditions, are presented in Figures \ref{fig:async_CNN_CIFAR10}, \ref{fig:async_cmp}, \ref{fig:async_cmp_cifar-10-IMDb}, and \ref{fig:async_cmp_cifar-100-FMNIST}. Generally, asynchronous model updates allow asynchronous methods to outpace their synchronous counterparts. The accuracy improvements observed in \AName{} and \SName{} stem from the dynamic adjustment of weights during the model aggregation process on both the server and devices. Additionally, the high training speed is attributed to the integration of the local model with the fresh global model during local training.

\subsubsection{Impact of Device Numbers}

\begin{figure*}[!t]
\centering
\subfigure[Asynchrorous]{
\includegraphics[width=0.6\linewidth]{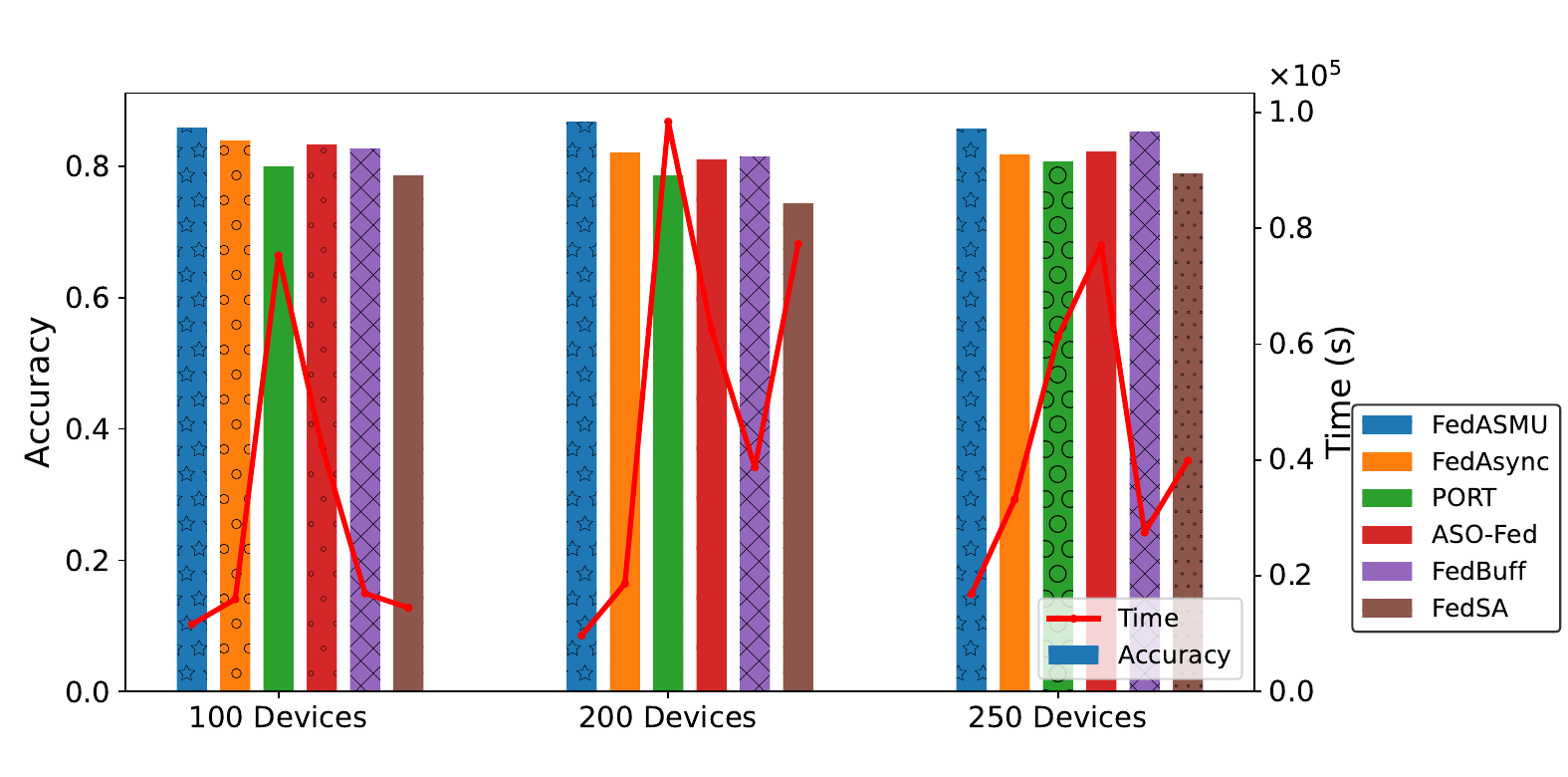}
}
\subfigure[Synchrorous]{
\includegraphics[width=0.6\linewidth]{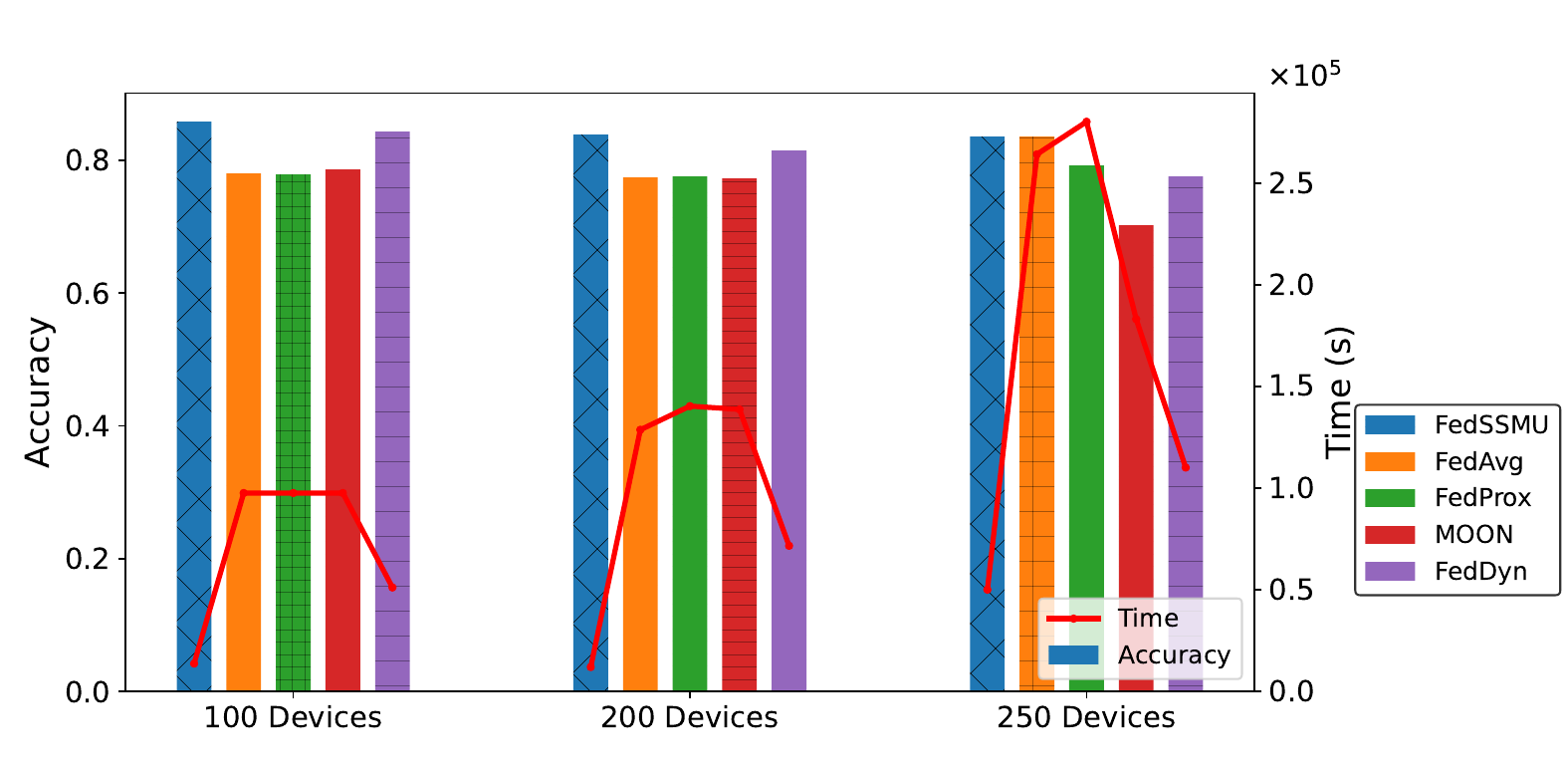}
}
\caption{Impact of device numbers.}
\label{fig:device_lenet_FMNIST}
\end{figure*}

\rev{
We conduct experiments with 100, 200 and 250 devices to demonstrate the remarkable scalability of \AName{} and \SName{}, as illustrated in Figure \ref{fig:device_lenet_FMNIST}. The performance of both methods significantly outpaces that of the baseline approaches, {achieving accuracy improvements ranging from 0.49\% to 12.51\% and speed enhancements of 19.83\% to 90.17\% for \AName{}. Similarly, \SName{} shows notable accuracy gains and improved efficiency (0.1\% to 13.38\% higher in accuracy and 54.5\% to 91.32\% faster)}. These results highlight the excellent scalability of both \AName{} and \SName{}.
}

\subsubsection{Impact of Device Heterogeneity}

\begin{figure*}[!t]
\centering
\subfigure[Asynchronous]{
\includegraphics[width=0.6\linewidth]{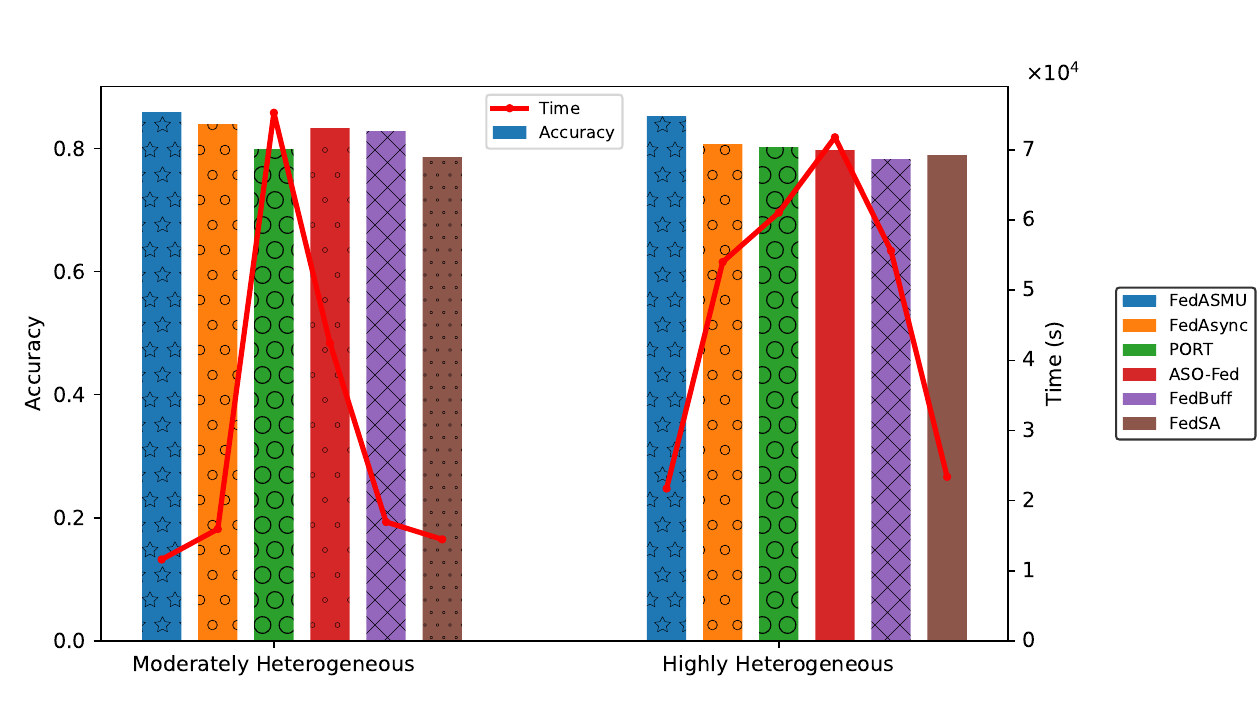}
}
\subfigure[Synchronous]{
\includegraphics[width=0.6\linewidth]{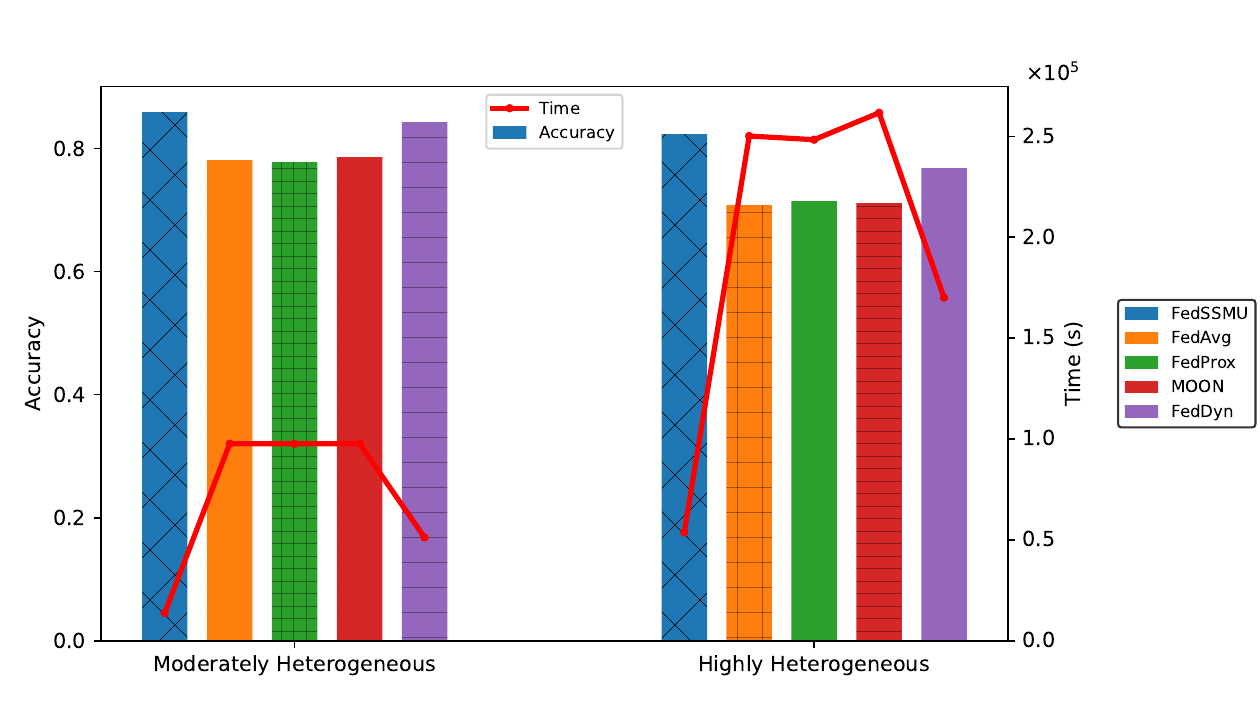}
}
\caption{Impact of device heterogeneity.}
\label{fig:heter_lenet_FMNIST}
\end{figure*}

We manipulate device heterogeneity regarding computation capacities to illustrate how \AName{} and \SName{} effectively address this variability, achieving remarkable accuracy and efficiency. We setup two levels for device heterogeneity. With the moderately heterogeneous level, the speed ratio between the fastest and slowest devices is 110, while the ratio is 440 with the highly heterogeneous level. The local training times for other devices randomly sampled within the maximum ratio for both levels. {As depicted in Figure \ref{fig:heter_lenet_FMNIST}, \AName{} demonstrates significant improvements, with accuracy increases ranging from 4.65\% to 7.07\% and efficiency gains of 7.12\% to 69.75\%. Similarly, \SName{} shows considerable performance enhancements, achieving accuracy boosts of 6.28\% to 12.38\% and efficiency increases of 68.37\% to 79.44\%.} When confronted with substantial differences among devices, both \AName{} and \SName{} effectively adapt model aggregation on both the server and devices, resulting in markedly better performance.

\subsubsection{Impact of Network Bandwidth}

\begin{figure*}[!t]
\centering
\subfigure[Asynchronous]{
\includegraphics[width=0.6\linewidth]{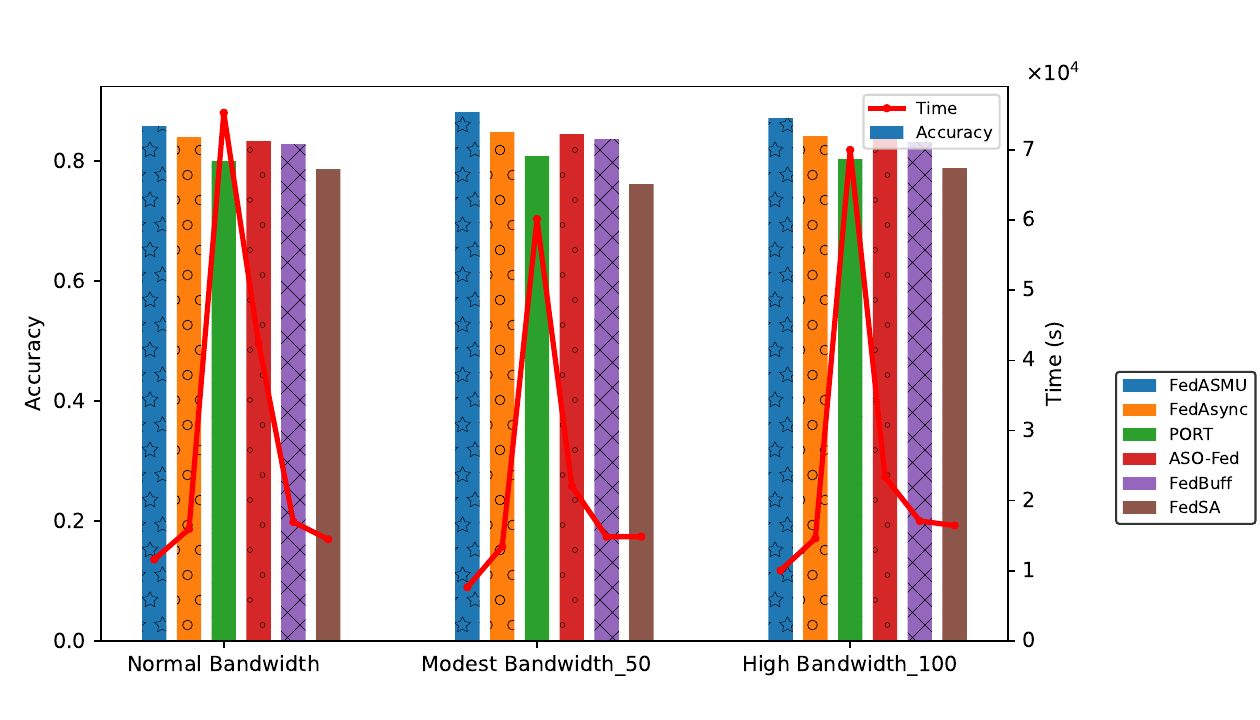}
}
\subfigure[Synchronous]{
\includegraphics[width=0.6\linewidth]{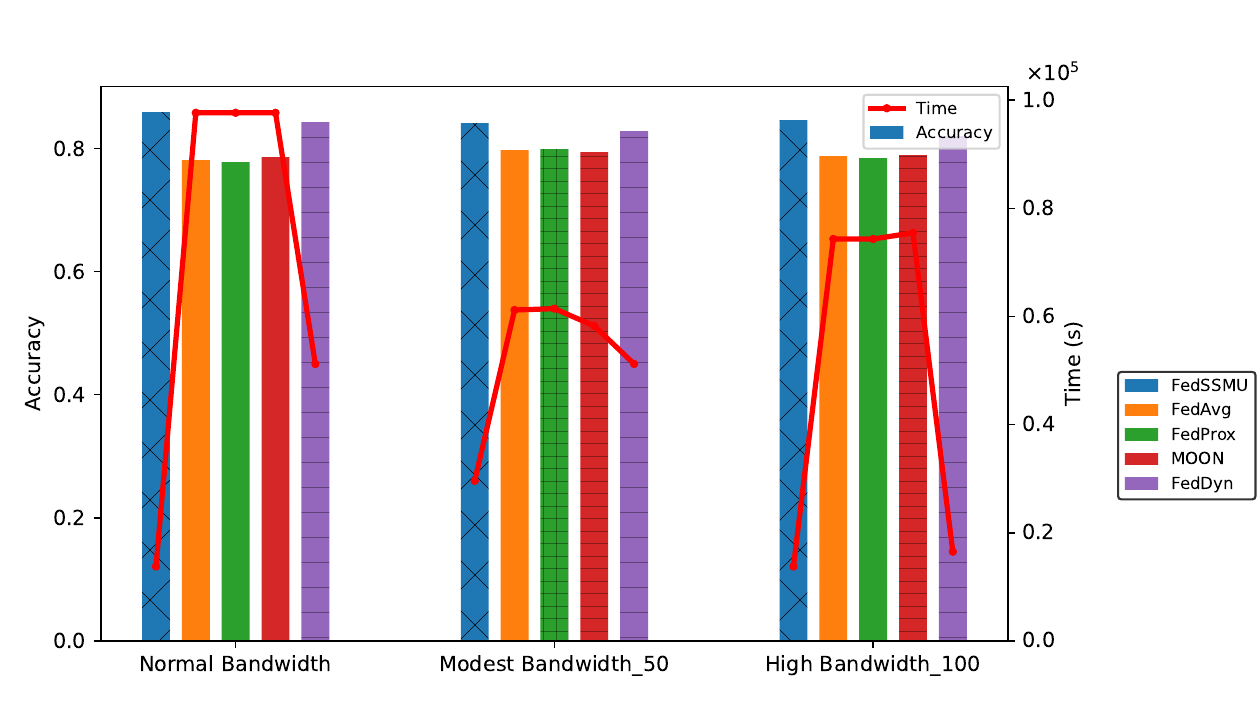}
}
\caption{Impact of network bandwidth.}
\label{fig:bandwidth_lenet_FMNIST}
\end{figure*}

As illustrated in Figure \ref{fig:bandwidth_lenet_FMNIST}, even with a significant reduction in network bandwidth (up to 50 times lower than normal), \AName{} continues to outperform baseline approaches,  achieving accuracy improvements ranging from 3.3\% to 11.93\% and efficiency gains between 43.1\% and 87.3\%. Similarly, \SName{} also demonstrates superior performance, with accuracy increases from 1.15\% to 4.58\% and efficiency enhancements from 42.2\% to 51.82\%. Although the advantages of both methods diminish due to the additional global model transfers, \AName{} and \SName{} still effectively enhance the efficiency of federated learning training, despite the increased data communication required to retrieve fresh global models.

\subsubsection{Communication Overhead Analysis}

\AName{} and \SName{} increase the communication overhead due to the additional transmissions of global models from server to devices. Since the downlink channel has high bandwidth, which incurs acceptable extra costs with significant benefits (higher accuracy and shorter training time). To analyze the performance of \AName{} and \SName{}, we carry out extra experimentation with the bandwidth of 100 (100 times smaller than normal). The advantages of \AName{} becomes even more significant compared with 50 (50 times smaller) (3.3\%-11.93\% for 50 to 2.97\%-{12.38}\% for 100) in terms of accuracy and (43.1\%-87.3\% for 50 to 31.28\%-{85.66}\% for 100) in terms of training time. ). The advantages of \SName{} becomes even more significant compared with 50 (50 times smaller) (1.15\%-4.58\% for 50 to 2.88\%-{6.1}\% for 100) in terms of accuracy and (42.2\%-51.82\% for 50 to 16.59\%-{81.78}\% for 100) in terms of training time. Such results reveal excellent performance of \AName{} and \SName{} within modest network environments.

\subsubsection{Ablation Study}

\begin{figure}[!t]
\centering
\subfigure[LeNet]{
\includegraphics[width=0.45\linewidth]{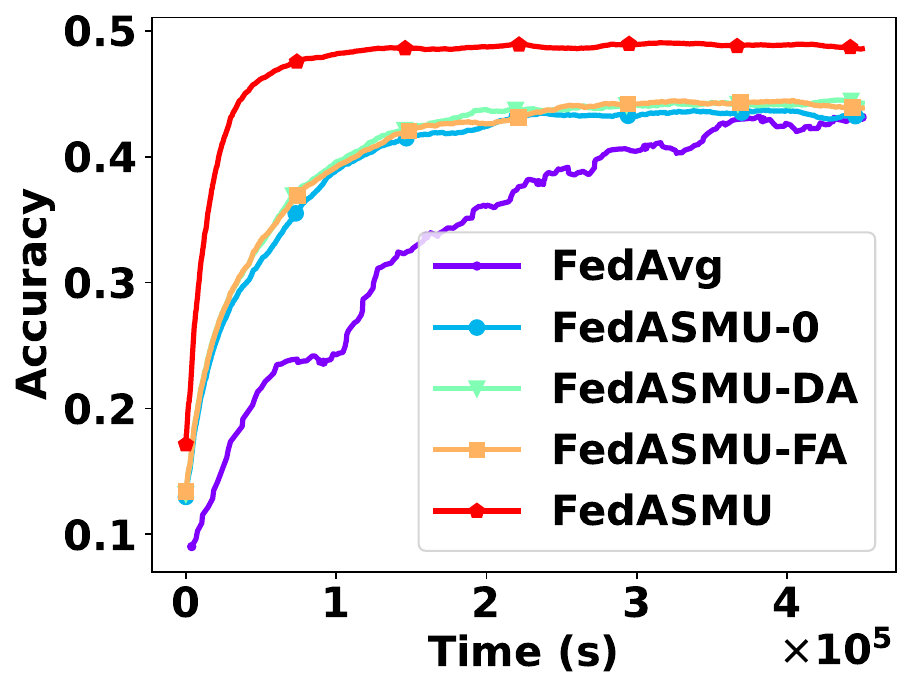}
\label{fig:abla_lenet_10}
}
\subfigure[CNN]{
\includegraphics[width=0.45\linewidth]{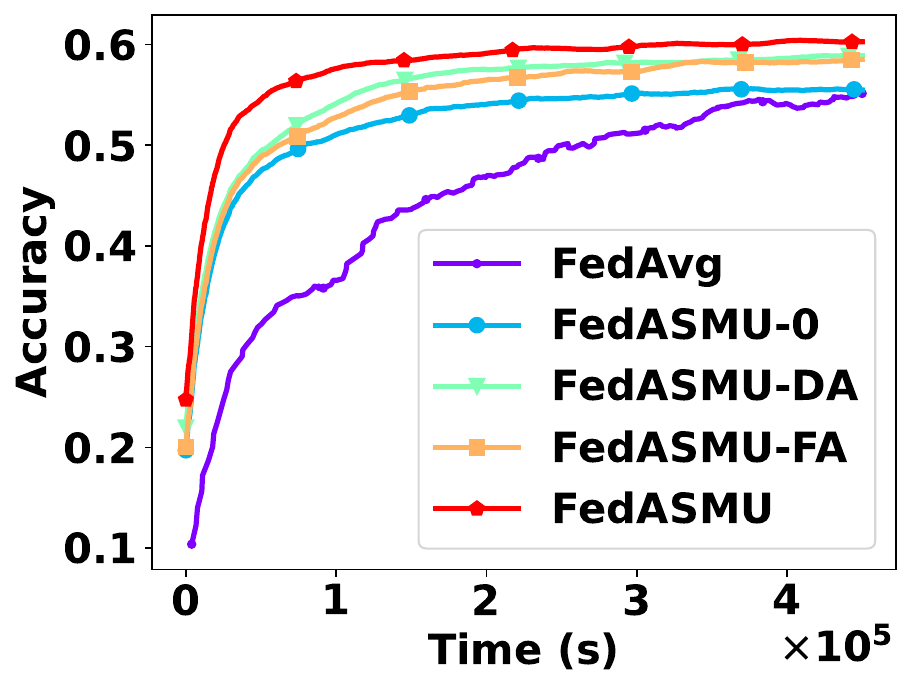}
\label{fig:abla_cnn_10}
}
\caption{The accuracy and training time with \AName{}, \AName{}-DA, \AName{}-FA, \AName{}-0, and FedAvg on CIFAR-10.}
\label{fig:ablation}
\end{figure}

As illustrated in Figure \ref{fig:ablation}, we perform an ablation study involving \AName{}-DA, \AName{}-FA, \AName{}-0, and FedAvg. In this context, \AName{}-DA indicates \AName{} without dynamic model aggregation on the server, while \AName{}-FA refers to \AName{} lacking adaptive model update on devices. \AName{}-0 denotes the version of \AName{} that excludes both methods, effectively making it comparable to FedAsync with a staleness bound.
The dynamic weight adjustment feature significantly enhances accuracy, allowing \AName{} to outperform \AName{}-DA by 1.38\% to 4.32\%. Similarly, \AName{}-FA surpasses \AName{}-0 with an accuracy improvement ranging from 0.65\% to 3.04\%. The adaptive model update on devices effectively minimizes the staleness between local and global models, enabling \AName{} to achieve target accuracy (0.30 for LeNet and 0.40 for CNN) with shorter training times—improving by 44.77\% to 73.96\% compared to \AName{}-FA, while also attaining higher accuracy gains of 1.75\% to 4.71\%.
Furthermore, \AName{}-DA demonstrates superior performance, with accuracy enhancements of 1.04\% to 3.41\% and improved efficiency, being 15.71\% to 19.54\% faster compared to \AName{}-0. Both \AName{}-DA and \AName{}-FA significantly outperform FedAvg in terms of accuracy, showing increases of 0.73\% to 3.75\%, as well as efficiency, which improves by 72.88\% to 85.72\%. Although \AName{}-0 exhibits slightly higher accuracy (0.08\% to 0.34\%) compared to FedAvg, it achieves substantially greater efficiency, being 67.84\% to 82.26\% faster due to its asynchronous mechanism.

To assess the effectiveness of the RL-based time slot selection method for devices requesting fresh global models, we compare it against three baseline strategies. The first hypothesis (H1) involves the device sending the request immediately after the first local epoch. The second hypothesis (H2) has the device sending the request midway through local training, while the third hypothesis (H3) dictates that the request is made in the penultimate local epoch. Our RL approach demonstrates a marked improvement in accuracy over these baselines, outperforming H1 by 2.4\%, H2 by 2.2\%, and H3 by 2.8\%.

\section{Conclusion}
\label{sec:con}

The delayed model update dissemination and the inefficient downlink bandwidth usage severerly limit the performance of FL.
In this paper, we propose a timely update dissemination approach for FL, which enables the server to promptly disseminate fresh model updates from devices that have completed one round of local training to other devices that are still conducting local training. 
Based on such an idea, in the asynchronous case, we propose a novel Asynchronous Staleness-aware Model Update FL framework, i.e., \AName{}, with an asynchronous system model and two novel methods, i.e., a dynamic model aggregation method on the server and an adaptive local model adjustment method on devices. We further extend the approach to the synchronous case, called \SName{}, which can also improve the performance of synchronous FL.
Extensive experimentation reveals significant advantages of \AName{} and \SName{} compared with synchronous and asynchronous baseline approaches in terms of accuracy 
(0.41\% to 145.87\% higher) and efficiency 
(3.73\% to 97.59\% faster).

\bibliographystyle{IEEEtran}
\bibliography{reference}

\appendix
\section{Appendix}
\label{sec:appendx}

\subsection{Convergence Analysis}

\begin{proof}
First, we denote the optimal model by $\boldsymbol{w}^*$, the new fresh global gradient is not received at the $l$-th local epoch, and we have the following inequality with the vanilla SGD in devices: 
\begin{align*}
&~\E\left[ \mathcal{F}(\boldsymbol{w}_{o,l}) - \mathcal{F}(\boldsymbol{w}^*) \right] \\
= &~\E_{\zeta_{l-1} \sim \mathcal{D}_i}\left[ \mathcal{F}(\boldsymbol{w}_{o,l-1} - \eta_i \nabla \mathcal{F}_{i}(\boldsymbol{w}_{o,l-1}, \zeta_{l-1})) - \mathcal{F}(\boldsymbol{w}^*) \right]\\
\leq &~\mathcal{F}(\boldsymbol{w}_{o,l-1}) - F(\boldsymbol{w}^*) \\
&\quad - \eta_i \E_{\zeta_{l-1} \sim \mathcal{D}_i}\left[  \ip{\nabla \mathcal{F}(\boldsymbol{w}_{o,l-1})}{\nabla \mathcal{F}_{i}(\boldsymbol{w}_{o,l-1}, \zeta_{l-1})} \right] \\
&\quad + \frac{L \eta_i^2}{2} \E_{\zeta_{l-1} \sim \mathcal{D}_i}\left[  \| \nabla \mathcal{F}_{i}(\boldsymbol{w}_{o,l-1}, \zeta_{l-1}) \|^2 \right]\\
\leq &~\mathcal{F}(\boldsymbol{w}_{o,l-1}) - F(\boldsymbol{w}^*)  + \frac{L \eta_i^2 \mathcal{G}^2}{2}\\
&\quad - \eta_i \underbrace{\E_{\zeta_{l-1} \sim \mathcal{D}_i}\left[  \ip{\nabla \mathcal{F}(\boldsymbol{w}_{o,l-1})}{\nabla \mathcal{F}_{i}(\boldsymbol{w}_{o,l-1}, \zeta_{l-1})} \right]}_{A}, \tag{1}\label{eq1}
\end{align*}
where the first inequality comes from $L$-smoothness and the second one is from bounded local gradient. Then, we focus on $A$. 
\begin{align*}
&~\mathbb{E}_{\zeta_{l-1} \sim \mathcal{D}_i} \parallel \nabla \mathcal{F}_i(\boldsymbol{w}_{o,l-1}; \zeta_{l-1}) - \nabla \mathcal{F}(\boldsymbol{w}_{o,l-1}) \parallel^2 \\
= &~\E\parallel \nabla \mathcal{F}(\boldsymbol{w}_{o,l-1}) \parallel^2  - 2A\\
&\quad + ~\mathbb{E}_{\zeta_{l-1} \sim \mathcal{D}_i} \parallel \nabla \mathcal{F}_i(\boldsymbol{w}_{o,l-1}; \zeta_{l-1}) \parallel^2. 
\end{align*}
Based on the bounded local variance assumption, we have:
\begin{align*}
\mathcal{V}^2
= ~&\E\parallel \nabla \mathcal{F}(\boldsymbol{w}_{o,l-1}) \parallel^2 - 2A \\
&\quad + ~\mathbb{E}_{\zeta_{l-1} \sim \mathcal{D}_i} \parallel \nabla \mathcal{F}_i(\boldsymbol{w}_{o,l-1}; \zeta_{l-1}) \parallel^2,
\end{align*}
and we can get $A$:
\begin{align*}
A = &\frac{1}{2} (\E\parallel \nabla \mathcal{F}(\boldsymbol{w}_{o,l-1}) \parallel^2 - ~\mathcal{V}^2 \\
&\quad + ~\mathbb{E}_{\zeta_{l-1} \sim \mathcal{D}_i} \parallel \nabla \mathcal{F}_i(\boldsymbol{w}_{o,l-1}; \zeta_{l-1}) \parallel^2),
\end{align*}
Plug this into Formula \ref{eq1}, and we have:
\begin{align*}
&~\E\left[ \mathcal{F}(\boldsymbol{w}_{o,l}) - \mathcal{F}(\boldsymbol{w}^*) \right] \\
\leq &~\mathcal{F}(\boldsymbol{w}_{o,l-1}) - F(\boldsymbol{w}^*) + \frac{L \eta_i^2 \mathcal{G}^2}{2}\\
&\quad - \frac{\eta_i}{2} (\E\parallel \nabla \mathcal{F}(\boldsymbol{w}_{o,l-1}) \parallel^2 - \mathcal{V}^2\\
&\quad + ~\mathbb{E}_{\zeta_{l-1} \sim \mathcal{D}_i} \parallel \nabla \mathcal{F}_i(\boldsymbol{w}_{o,l-1}; \zeta_{l-1}) \parallel^2 ) \\
\leq &~\mathcal{F}(\boldsymbol{w}_{o,l-1}) - F(\boldsymbol{w}^*) - \frac{\eta_i}{2} \E\parallel \nabla \mathcal{F}(\boldsymbol{w}_{o,l-1}) \parallel^2 \\
&\quad + \frac{L \eta_i^2 \mathcal{G}^2 + \eta_i \mathcal{V}^2}{2},
\end{align*}
where the second inequality is because $\mathbb{E}_{\zeta_{l-1} \sim \mathcal{D}_i} \parallel \nabla \mathcal{F}_i(\boldsymbol{w}_{o,l-1}; \zeta_{l-1}) \parallel^2 \geq 0$. By rearranging the terms and telescoping, we have:
\begin{align*}
\E\parallel \nabla \mathcal{F}(\boldsymbol{w}_{o,l-1}) \parallel^2 \leq &\frac{2}{\eta_i} \E\left[ \mathcal{F}(\boldsymbol{w}_{o,l-1}) - \mathcal{F}(\boldsymbol{w}_{o,l}) \right] \\
&\quad + L \eta_i \mathcal{G}^2 + \mathcal{V}^2.
\end{align*}
However, when the fresh global model $\boldsymbol{w}_g$ is received right at the $l^*$-th local epoch, we have:
\begin{align*}
&\E\parallel \nabla \mathcal{F}(\boldsymbol{w}_{o,l^*}) \parallel^2 \\
\leq &\frac{2}{\eta_i} \E\left[ \mathcal{F}(\boldsymbol{w}_{o,l^*-1}^a) - \mathcal{F}(\boldsymbol{w}_{o,l^*}) \right] + L \eta_i \mathcal{G}^2 + \mathcal{V}^2 \\
= &\frac{2}{\eta_i} \E\left[ \mathcal{F}((1-\beta^i_{t_i-1}) \boldsymbol{w}_{o,l^*-1}^b + \beta^i_{t_i-1} \boldsymbol{w}_g) - \mathcal{F}(\boldsymbol{w}_{o,l^*}) \right] \\
&\quad + L \eta_i \mathcal{G}^2 + \mathcal{V}^2 \\
\leq &\frac{2}{\eta_i} \E\left[ (1-\beta^i_{t_i-1}) \mathcal{F}(\boldsymbol{w}_{o,l^*-1}^b) + \beta^i_{t_i-1} \mathcal{F}(\boldsymbol{w}_g) -  \mathcal{F}(\boldsymbol{w}_{o,l^*}) \right] \\
&\quad + L \eta_i \mathcal{G}^2 + \mathcal{V}^2 \\
= &\frac{2}{\eta_i} \E\left[ (1-\beta^i_{t_i-1}) \mathcal{F}(\boldsymbol{w}_{o,l^*-1}) + \beta^i_{t_i-1} \mathcal{F}(\boldsymbol{w}_g) - \mathcal{F}(\boldsymbol{w}_{o,l^*}) \right] \\
&\quad + L \eta_i \mathcal{G}^2 + \mathcal{V}^2 \\
= &\frac{2}{\eta_i} \E\left[ \mathcal{F}(\boldsymbol{w}_{o,l^*-1}) - \mathcal{F}(\boldsymbol{w}_{o,l^*}) \right]\\
&\quad + \frac{2 \beta^i_{t_i-1}}{\eta_i} \E\left[ \mathcal{F}(\boldsymbol{w}_g) -\mathcal{F}(\boldsymbol{w}_{o,l^*-1}) \right] \\
&\quad + L \eta_i \mathcal{G}^2 + \mathcal{V}^2 \tag{2}\label{eq2}
\end{align*}
where the second inequality is because of convexity of $\mathcal{F}(\cdot)$. Then, we can get:
\begin{align*}
&~\sum_{l=1}^{\mathcal{L}^i} \E\parallel \nabla \mathcal{F}(\boldsymbol{w}_{o,l}) \parallel^2 \\
\leq &~\frac{2}{\eta_i} \E\left[ \mathcal{F}(\boldsymbol{w}_{o,0}) - \mathcal{F}(\boldsymbol{w}_{o,\mathcal{L}^i}) \right] + \mathcal{L}^i (L \eta_i \mathcal{G}^2 + \mathcal{V}^2)\\
&\quad + \frac{2 \beta^i_{t_i-1}}{\eta_i} \E\left[ \mathcal{F}(\boldsymbol{w}_g) -\mathcal{F}(\boldsymbol{w}_{o,l^*-1}) \right] \\
= &~\frac{2}{\eta_i} \E\left[ \mathcal{F}(\boldsymbol{w}_o) - \mathcal{F}(\boldsymbol{w}_o^i)  \right] + \mathcal{L}^i (L \eta_i \mathcal{G}^2 + \mathcal{V}^2)\\
&\quad + \frac{2 \beta^i_{t_i-1}}{\eta_i} \E\left[ \mathcal{F}(\boldsymbol{w}_g) -\mathcal{F}(\boldsymbol{w}_{o,l^*-1}) \right] \\
= &~\frac{2}{\eta_i} \underbrace{\E\left[ \mathcal{F}(\boldsymbol{w}_o) - \mathcal{F}(\boldsymbol{w}_o^i)\right]}_{B} + \mathcal{L}^i (L \eta_i \mathcal{G}^2 + \mathcal{V}^2)\\
&\quad + \frac{2}{\eta_i} \beta^i_{t_i-1}\underbrace{\E\left[ \mathcal{F}(\boldsymbol{w}_{o,0}) - \mathcal{F}(\boldsymbol{w}_{o,l^*-1}) \right]}_{C} \\
&\quad + \frac{2}{\eta_i} \beta^i_{t_i-1}\underbrace{\E\left[ \mathcal{F}(\boldsymbol{w}_g) - \mathcal{F}(\boldsymbol{w}_o) \right]}_{D}.
\end{align*}
First, we focus on the calculation of $B$. 
\begin{align*}
&\E\left[ \mathcal{F}(\boldsymbol{w}_{t+1}) - \mathcal{F}(\boldsymbol{w}_{t}) \right] \\
= &~\E\left[ \mathcal{F}((1-\alpha^i_t) \boldsymbol{w}_{t} + \alpha^i_t \boldsymbol{w}_o^i) - \mathcal{F}(\boldsymbol{w}_{t}) \right] \\
\leq &~\E\left[ (1-\alpha^i_t) \mathcal{F}(\boldsymbol{w}_{t}) + \alpha^i_t \mathcal{F}(\boldsymbol{w}_o^i) - \mathcal{F}(\boldsymbol{w}_{t}) \right] \\
= &~\alpha^i_t \E\left[ \mathcal{F}(\boldsymbol{w}_o^i)  - \mathcal{F}(\boldsymbol{w}_{t}) \right] \\
= &~\alpha^i_t \E\left[ \mathcal{F}(\boldsymbol{w}_o^i) - \mathcal{F}(\boldsymbol{w}_o) + \mathcal{F}(\boldsymbol{w}_o)  - \mathcal{F}(\boldsymbol{w}_{t}) \right],
\end{align*}
where the inequility is because $\mathcal{F}(\cdot)$ is convex. Then, we have:
\begin{align*}
&~\E\left[ \mathcal{F}(\boldsymbol{w}_{t+1}) - \mathcal{F}(\boldsymbol{w}_{t}) \right] \\
\leq &~\alpha^i_t \E\left[ \mathcal{F}(\boldsymbol{w}_o^i) - \mathcal{F}(\boldsymbol{w}_o) + \mathcal{F}(\boldsymbol{w}_o)  - \mathcal{F}(\boldsymbol{w}_{t}) \right].
\end{align*}
And, we can get: 
\begin{align*}
&~\E\left[ \mathcal{F}(\boldsymbol{w}_o) - \mathcal{F}(\boldsymbol{w}_o^i) \right] \\
\leq &~\frac{1}{\alpha^i_t} \E\left[ \mathcal{F}(\boldsymbol{w}_{t}) - \mathcal{F}(\boldsymbol{w}_{t+1}) \right] + \E\left[ \mathcal{F}(\boldsymbol{w}_o)  - \mathcal{F}(\boldsymbol{w}_{t}) \right] .
\end{align*}
Using $L$-smoothness, we have:
\begin{align*}
&~\E\left[ \mathcal{F}(\boldsymbol{w}_o)  - \mathcal{F}(\boldsymbol{w}_t) \right] \\
\leq &~\ip{\nabla \mathcal{F}(\boldsymbol{w}_t)}{\boldsymbol{w}_o - \boldsymbol{w}_t} + \frac{L}{2}\parallel \boldsymbol{w}_o - \boldsymbol{w}_t \parallel ^2 \\
\leq &~\parallel \nabla \mathcal{F}(\boldsymbol{w}_t) \parallel \parallel \boldsymbol{w}_o - \boldsymbol{w}_t \parallel + \frac{L}{2}\parallel \boldsymbol{w}_o - \boldsymbol{w}_t \parallel ^2
\end{align*}
As the fresh global model is incurred to reduce the difference between the local model and the global model, the difference between the global models of two versions is because of the local updates. Then, we have the upper bound of local updates:
\begin{align*}
&~\parallel \boldsymbol{w}_{o,0}  - \boldsymbol{w}_{o,\mathcal{L}^i} \parallel \\
\leq &~\parallel \boldsymbol{w}_{o,0}  - \boldsymbol{w}_{o,1} \parallel + \parallel \boldsymbol{w}_{o,1}  - \boldsymbol{w}_{o,2} \parallel + \cdots \\
& \quad + \parallel \boldsymbol{w}_{o,\mathcal{L}^i - 1}  - \boldsymbol{w}_{o,\mathcal{L}^i} \parallel \\
\leq &~\eta_i \mathcal{L}^i \mathcal{G}.
\end{align*}
And, we get:
\begin{align*}
\parallel \boldsymbol{w}_{o} - \boldsymbol{w}_{o+1} \parallel &= \parallel \boldsymbol{w}_{o} - (1-\alpha_t^i) \boldsymbol{w}_{o} - \alpha_t^i \boldsymbol{w}_{o,\mathcal{L}^i} \parallel \\
&= \alpha_t^i \parallel \boldsymbol{w}_{o,0} - \boldsymbol{w}_{o,\mathcal{L}^i} \parallel \\
&\leq \eta_i \alpha_t^i \mathcal{L} \mathcal{G}.
\end{align*}
Thus, we have:
\begin{align*}
\parallel \boldsymbol{w}_o - \boldsymbol{w}_t \parallel &\leq (t - o + 1) \eta_i \alpha_t^i \mathcal{L}^i \mathcal{G},
\end{align*}
where $t - o + 1\leq \tau$ because of staleness bound. Then, we can get:
\begin{align*}
\parallel \boldsymbol{w}_o - \boldsymbol{w}_t \parallel &\leq \tau \eta_i \alpha_t^i \mathcal{L}^i \mathcal{G}.
\end{align*}
Then, we have:
\begin{align*}
&~\E\left[ \mathcal{F}(\boldsymbol{w}_o)  - \mathcal{F}(\boldsymbol{w}_t) \right] \\
\leq &~\parallel \nabla \mathcal{F}(\boldsymbol{w}_t) \parallel \parallel \boldsymbol{w}_o - \boldsymbol{w}_t \parallel + \frac{L}{2}\parallel \boldsymbol{w}_o - \boldsymbol{w}_t \parallel ^2 \\
&\leq \tau \eta_i \alpha_t^i \mathcal{L}^i \mathcal{G}^2 + \frac{L}{2} (\tau \eta_i \alpha_t^i \mathcal{L}^i \mathcal{G})^2.
\end{align*}
And, we can calculate $B$:
\begin{align*}
B \leq &\frac{1}{\alpha^i_t} \E\left[ \mathcal{F}(\boldsymbol{w}_{t}) - \mathcal{F}(\boldsymbol{w}_{t+1}) \right] + \tau \eta_i \alpha_t^i \mathcal{L} \mathcal{G}^2 \\
&\quad + \frac{L}{2} (\tau \eta_i \alpha_t^i \mathcal{L} \mathcal{G})^2.
\end{align*}
Now, we focus on the calculation of $C$. Based on the convexity of $\mathcal{F}(\cdot)$, we have:
\begin{align*}
&~\E\left[ \mathcal{F}(\boldsymbol{w}_{o,l-1}) - \mathcal{F}(\boldsymbol{w}_{o,l}) \right] \\
\leq &~\ip{\nabla \mathcal{F}(\boldsymbol{w}_{o,l})}{\boldsymbol{w}_{o,l-1} - \boldsymbol{w}_{o,l}} + \frac{L}{2}\parallel \boldsymbol{w}_{o,l-1} - \boldsymbol{w}_{o,l} \parallel ^2 \\
= &~\eta_i \ip{\nabla \mathcal{F}(\boldsymbol{w}_{o,l})}{\nabla \mathcal{F}_i(\boldsymbol{w}_{o,l-1})} + \frac{L\eta_i^2}{2}\parallel \nabla \mathcal{F}_i(\boldsymbol{w}_{o,l-1}) \parallel ^2 \\
\leq &~\frac{\eta_i}{2} (\parallel \nabla \mathcal{F}(\boldsymbol{w}_{o,l}) \parallel^2 + \parallel \nabla \mathcal{F}_i(\boldsymbol{w}_{o,l-1}) \parallel^2) + \frac{L \eta_i^2 \mathcal{G}^2}{2} \\
\leq &~ \frac{2 \eta_i  + L \eta_i^2}{2} \mathcal{G}^2,
\end{align*}
Then, we can have:
\begin{align*}
C = &~\E\left[ \mathcal{F}(\boldsymbol{w}_{o,0}) - \mathcal{F}(\boldsymbol{w}_{o,l^*-1}) \right] \\
\leq &~\frac{2 \eta_i  + L \eta_i^2}{2} (l^*-1) \mathcal{G}^2 \\
\leq &~\frac{2 \eta_i  + L \eta_i^2}{2} \mathcal{L}^i \mathcal{G}^2.
\end{align*}
Next, we focus on the calculation of $D$. 
\begin{align*}
D &\leq \ip{\nabla \mathcal{F}(\boldsymbol{w}_o)}{\boldsymbol{w}_g - \boldsymbol{w}_o} + \frac{L}{2} \|\boldsymbol{w}_g - \boldsymbol{w}_o\|^2 \\
&\leq \|\nabla \mathcal{F}(\boldsymbol{w}_o)\| \|\boldsymbol{w}_g - \boldsymbol{w}_o\| + \frac{L}{2} \|\boldsymbol{w}_g - \boldsymbol{w}_o\|^2. 
\end{align*}
As $(g - o \leq \tau)$ because of staleness bound, we have:
\begin{align*}
\parallel \boldsymbol{w}_g - \boldsymbol{w}_o \parallel \leq (g - o) \eta_i \alpha_t^i \mathcal{L} \mathcal{G} \leq \tau \eta_i \alpha_t^i \mathcal{L} \mathcal{G}.
\end{align*}
Then, we have:
\begin{align*}
D &\leq \|\nabla \mathcal{F}(\boldsymbol{w}_o)\| \|\boldsymbol{w}_g - \boldsymbol{w}_o\| + \frac{L}{2} \|\boldsymbol{w}_g - \boldsymbol{w}_o\|^2 \\
&\leq \tau \eta_i \alpha_t^i \mathcal{L} \mathcal{G}^2 + \frac{L}{2} (\tau \eta_i \alpha_t^i \mathcal{L} \mathcal{G})^2.
\end{align*}
By rearranging the terms, we have
\begin{align*}
&~\sum_{l=1}^{\mathcal{L}^i} \E\parallel \nabla \mathcal{F}(\boldsymbol{w}_{o,l}) \parallel^2 \\
\leq &~\frac{2}{\eta_i} \underbrace{\E\left[ \mathcal{F}(\boldsymbol{w}_o) - \mathcal{F}(\boldsymbol{w}_o^i)\right]}_{B} \\
&\quad + \frac{2}{\eta_i} \beta^i_{t_i-1}\underbrace{\E\left[ \mathcal{F}(\boldsymbol{w}_{o,0}) - \mathcal{F}(\boldsymbol{w}_{o,l^*-1}) \right]}_{C} \\
&\quad + \frac{2}{\eta_i} \beta^i_{t_i-1}\underbrace{\E\left[ \mathcal{F}(\boldsymbol{w}_g) - \mathcal{F}(\boldsymbol{w}_o) \right]}_{D} + \mathcal{L}^i (L \eta_i \mathcal{G}^2 + \mathcal{V}^2) \\
\leq & ~\frac{2}{\eta_i} (\frac{1}{\alpha^i_t} \E\left[ \mathcal{F}(\boldsymbol{w}_{t}) - \mathcal{F}(\boldsymbol{w}_{t+1}) \right]) \\
&\quad + \frac{2}{\eta_i} (\tau \eta_i \alpha_t^i \mathcal{L}^i \mathcal{G}^2 + \frac{L}{2} (\tau \eta_i \alpha_t^i \mathcal{L}^i \mathcal{G})^2) \\
&\quad + \beta^i_{t_i-1} \frac{2 \eta_i  + L \eta_i^2}{2} \mathcal{L}^i \mathcal{G}^2 \\
&\quad + \beta^i_{t_i-1} (\tau \eta_i \alpha_t^i \mathcal{L}^i \mathcal{G}^2 + \frac{L}{2} (\tau \eta_i \alpha_t^i \mathcal{L}^i \mathcal{G})^2) \\
&\quad + \mathcal{L}^i (L \eta_i \mathcal{G}^2 + \mathcal{V}^2) \\
= & ~\frac{2 \E\left[ \mathcal{F}(\boldsymbol{w}_{t}) - \mathcal{F}(\boldsymbol{w}_{t+1}) \right]}{\alpha^i_t \eta_i} + \mathcal{L}^i \mathcal{V}^2 \\
&\quad + \frac{\beta^i_{t_i-1} + \tau^2 (\alpha_t^i)^2 \mathcal{L}^i}{2} L \mathcal{L}^i \eta_i^2 \mathcal{G}^2\\
&\quad + (2 \tau \alpha_t^i + L \tau^2  \eta_i (\alpha_t^i)^2 \mathcal{L}^i + \beta^i_{t_i-1} \eta_i \\
&\quad + \beta^i_{t_i-1} \tau + \eta_i \alpha_t^i + L \eta_i) \mathcal{L}^i \mathcal{G}^2.
\end{align*}
We take $\alpha_{min} \leq \alpha_t^i \leq 1$ and $0 \leq \beta^i_{t_i-1} \leq 1$ with $\alpha_{min} > 0$, we can get:
\begin{align*}
&~\sum_{l=1}^{\mathcal{L}^i} \E\parallel \nabla \mathcal{F}(\boldsymbol{w}_{o,l}) \parallel^2 \\
\leq & ~\frac{2 \E\left[ \mathcal{F}(\boldsymbol{w}_{t}) - \mathcal{F}(\boldsymbol{w}_{t+1}) \right]}{\alpha_{min} \eta_i} + \mathcal{L}^i \mathcal{V}^2 + \frac{1 + \tau^2 \mathcal{L}^i}{2} L \mathcal{L}^i \eta_i^2 \mathcal{G}^2\\
& + (3 \tau + L \tau^2  \eta_i \mathcal{L}^i + 2 \eta_i + L \eta_i) \mathcal{L}^i \mathcal{G}^2.
\end{align*}
After $T$ global rounds, we have:
\begin{align*}
&~\frac{1}{\sum_{t=0}^{T}\mathcal{L}_t} \sum_{t=0}^{T} \sum_{l=0}^{\mathcal{L}_t}\E\parallel \nabla \mathcal{F}(\boldsymbol{w}_{o,l}) \parallel^2 \\
\leq &~\frac{2 \E\left[ \mathcal{F}(\boldsymbol{w}_{0}) - \mathcal{F}(\boldsymbol{w}_{T}) \right]}{\alpha_{min} \eta_i T \mathcal{L}_{min}} + \frac{1 + \tau^2 \mathcal{L}_{max}}{2T \mathcal{L}_{min}} L \mathcal{L}_{max} \eta_i \mathcal{G}^2\\
& + \frac{3 \tau + L \tau^2  \eta_i \mathcal{L}_{max} + 2 \eta_i + L \eta_i}{T \mathcal{L}_{min}}\mathcal{L}_{max} \mathcal{G}^2 + \frac{\mathcal{L}_{max} \mathcal{V}^2}{T \mathcal{L}_{min}},
\end{align*}
where $\mathcal{L}_t$ represents the maximum local epochs within the $t$-th global round with $\mathcal{L}_{min} \leq \mathcal{L}_t \leq \mathcal{L}_{max}$. We take  $\eta_i = \frac{1}{\sqrt{T}}$ and $T = \mathcal{L}_{min}^6$, and can get:
\begin{align*}
&~\frac{1}{\sum_{t=0}^{T}\mathcal{L}_t} \sum_{t=0}^{T} \sum_{l=0}^{\mathcal{L}_t}\E\parallel \nabla \mathcal{F}(\boldsymbol{w}_{o,l}) \parallel^2 \\
\leq & ~\frac{2 \E\left[ \mathcal{F}(\boldsymbol{w}_{0}) - \mathcal{F}(\boldsymbol{w}_{T}) \right]}{\alpha_{min} \mathcal{L}_{min}^3} + \frac{1 + \tau^2 \mathcal{L}_{max}}{2 \mathcal{L}_{min}^3} L \mathcal{L}_{max} \mathcal{G}^2\\
&\quad + \frac{3 \tau \mathcal{L}_{max} \mathcal{G}^2 + \mathcal{L}^i \mathcal{V}^2}{ \mathcal{L}_{min}^7} + \frac{ L \tau^2 \mathcal{L}_{max} + 2 + L }{\mathcal{L}_{min}^3}\mathcal{L}_{max} \mathcal{G}^2 \\
\leq & \frac{2 \E\left[ \mathcal{F}(\boldsymbol{w}_{0}) - \mathcal{F}(\boldsymbol{w}_{T}) \right]}{\alpha_{min} \mathcal{L}_{min}^3} + \OM\left( \frac{L \mathcal{G}^2 \mathcal{L}_{max}}{\mathcal{L}_{min}^3} \right) \\
&\quad + \OM\left( \frac{\mathcal{L}^i \mathcal{V}^2}{ \mathcal{L}_{min}^7}\right) + \OM\left( \frac{\tau \mathcal{G}^2 \mathcal{L}_{max}}{ \mathcal{L}_{min}^7} \right) \\
&\quad + \OM\left( \frac{ \mathcal{G}^2 \mathcal{L}_{max} }{\mathcal{L}_{min}^3} \right) + \OM\left( \frac{ L \mathcal{G}^2 \mathcal{L}_{max} }{\mathcal{L}_{min}^3} \right)\\
&  + \OM\left(\frac{ L \tau^2 \mathcal{G}^2 \mathcal{L}_{max}^2}{\mathcal{L}_{min}^3} \right) + \OM\left( \frac{ L \tau^2 \mathcal{G}^2 \mathcal{L}_{max}^2 }{\mathcal{L}_{min}^3} \right) 
\end{align*}

\end{proof}

\end{document}